\documentclass[fleqn, usenatbib]{mnras}

\usepackage[T1]{fontenc}

\DeclareRobustCommand{\VAN}[3]{#2}
\let\VANthebibliography\thebibliography
\def\thebibliography{\DeclareRobustCommand{\VAN}[3]{##3}\VANthebibliography}

\usepackage{booktabs}
\usepackage{pgfplotstable}
\pgfplotsset{compat=1.18}
\pgfplotstableset{
precision=4,
}
\usepackage{siunitx}
\usepackage{graphicx}
\usepackage{amsmath}
\usepackage{amssymb}
\usepackage{subcaption}
\usepackage{hyperref}
\usepackage{lipsum}
\usepackage{pdflscape}
\usepackage{xcolor}
\usepackage{anyfontsize}

\newcommand{\msol}{$\mathrm{M_{\sun}}$}

\newcommand{\cm}{$\mathrm{cm}$}

\newcommand{\K}{$\mathrm{K}$}
\newcommand{\s}{$\mathrm{s}$}

\newcommand{\gcmcubed}{$\mathrm{g}\,\mathrm{cm}^{-3}$}
\newcommand{\kmpersec}{$\mathrm{km}\,\mathrm{s}^{-1}$}

\title[Ne-22 core and shell WD simulations]{Comparison of Ne-22 core and shell distilled WD detonations in AREPO} 

\author[U. P. Burmester et al.]{
Uri Pierre Burmester$^{1}$\thanks{E-mail: uri.burmester@anu.edu.au (UPB)},
Lilia Ferrario$^{1}$,
Ivo R.~Seitenzahl$^{1,2}$, and
Simon Blouin$^{3}$
\\
$^{1}$Mathematical Sciences Institute, Australian National University, Canberra ACT 0200, AU\\
$^{2}$Research School of Astronomy and Astrophysics, Australian National University, Canberra ACT 2611, Australia\\
${3}$Department of Physics and Astronomy, University of Victoria, Victoria, British Columbia, Canada
}

\date{Accepted XXX. Received YYY; in original form ZZZ}

\pubyear{2025}

\begin{document}
\label{firstpage}
\pagerange{\pageref{firstpage}--\pageref{lastpage}}
\maketitle

\begin{abstract}
We present three-dimensional hydrodynamical simulations of detonations in $1.0$\,\msol white dwarfs that have undergone $^{22} \mathrm{Ne}$ distillation during crystallisation. These simulations, conducted with the moving-mesh code \textsc{AREPO}, aim to investigate the effects of chemical separation on the ejecta and spectra of such WDs undergoing thermonuclear explosions. The distillation process alters the internal chemical stratification of the star, concentrating neutron-rich material either in a central core or in an interior shell. We model both configurations as well as a homogeneous equivalent for each case with the same $^{22} \mathrm{Ne}$ content distributed evenly at all radii. Despite similar $^{56} \mathrm{Ni}$ yields between the core and shell models ($0.40$ and $0.45$\,\msol\ respectively), the two models yield markedly different iron-group abundances. Both distilled models showed significantly enhanced production of $^{15} \mathrm{N}$ via the decay of $^{15} \mathrm{O}$. The $^{22} \mathrm{Ne}$-core model produces enhanced amounts of stable neutron-rich iron-group isotopes such as $^{58} \mathrm{Ni}$ and $^{54} \mathrm{Fe}$. We highlight observational signatures associated with these differences, including potentially enhanced [\ion{Ni}{II}] lines in nebular spectra. Synthetic \textsc{TARDIS} spectra at early times show only moderate differences. Our results suggest that white dwarf distillation, a process linked to delayed cooling in the Gaia Q branch population, may leave detectable nucleosynthetic fingerprints in a subset of Type Ia supernovae. These findings open additional pathways to probe progenitor evolution and the role of crystallisation in shaping the diversity of thermonuclear transients.
\end{abstract}




\section{Introduction} \label{sec:introduction}

The process of WD cooling provides a useful ``cosmic clock'' that can be used for the purposes of dating stellar populations. Due to the predictable thermal evolution of WDs, their effective temperature serves as a proxy for the star's age, enabling age-dating of nearby stellar populations, the reconstruction of star formation histories, and even insights into the composition of rocky exoplanets that once orbited them \citep[e.g.,][]{wingetIndependentMethodDetermining1987,isernStarFormationHistory2019,hollandsCoolDZWhite2017,hollandsCoolDZWhite2018}. Typically, WD ages are inferred by mapping spectroscopic or photometric measurements of their effective temperatures and masses onto theoretical cooling tracks. However, these estimates generally account only for observational uncertainties, neglecting systematic effects inherent in the cooling models themselves.

Our understanding of the WD cooling process has been significantly enhanced in recent years, particularly due to the vast increase in the number of objects available for theoretical study. In particular, the second data release from the Gaia collaboration (DR2) revealed three individual cooling branches, dubbed the A, B, and Q branches \citep{brownGaiaDataRelease2018}. The A and B cooling tracks are dominated by WDs with hydrogen- and helium-rich atmospheres \citep{chengCoolingAnomalyHighmass2019}. The Q branch, however, is not aligned with any cooling track or isochrone. It has been suggested that this branch is caused by a delay of cooling instead of a peak in mass or age distribution, resulting in a Q branch overdensity in the Gaia colour-magnitude diagram \citep{tremblayCoreCrystallizationPileup2019}.

\citet{tremblayCoreCrystallizationPileup2019} also identified the Q branch as corresponding to the predicted location of ultramassive ($\gtrsim 1 \,M_{\odot}$) WDs undergoing crystallisation -- a transition that occurs during the WD cooling process by which the star reaches a critical temperature and undergoes a liquid-to-solid phase transition \citep{tremblayGaiaWhiteDwarf2024}. This phase transition releases gravitational energy due to chemical separation as well as latent heat, which is predicted to result in a cooling delay of $1 - 2$\,Gyr \citep{bauerCarbonOxygenPhase2023}. However, these sources of energy are not sufficient to explain the incidence of WDs appearing in the Q branch cooling delay as they crystallise \citep{blouinPrecisionCosmochronologyNew2020}. The phase separation of $^{22} \mathrm{Ne}$-rich liquid around a growing crystallised core has been proposed to explain the additional cooling delay. This process is called ``distillation''. 

Understanding the effects of distillation on a WD's observable features is vital for accurate categorisation of observations. Some of the relevant questions include: whether distilled WDs that explode as type Ia supernovae might appear distinct when viewed from Earth, and what effect does distillation have on the supernova ejecta. This may be useful, for example, in confirming the mechanism of the additional cooling delay observed in some members of the Q branch. In this work, we consider the detonation and resulting observables of two distilled WD models using the moving-mesh code \textsc{AREPO} and the radiative transfer code \textsc{TARDIS}. We also consider two homogeneous WDs with the same overall composition as the distilled WDs, to isolate the effect of the separation of chemical species on the ejecta. The relevant literature is summarised in Section~\ref{sec:litreview}. The codes and the parameters used to set up our simulations are described in Section~\ref{sec:methods}. Section~\ref{sec:results} documents our findings. We first examine the stability of the WD structures in Section~\ref{sec:results_relaxation}, after which we discuss the properties of the ejecta and spectra generated by our codes. A summary is provided in Section~\ref{sec:conclusion}. 

\section{Prior Work} \label{sec:litreview}

WD crystallisation is a relatively well-understood process that arises as a result of the cooling experienced by WD populations as they age. As first described by \citet{vanhornCrystallizationWhiteDwarfs1968}, cooling WDs have an initially high temperature that gradually decreases due to the loss of energy via radiation. The motion of the plasma interior to the WD is governed by the ratio of the Coulomb to thermal energy. As the WD cools, this ratio increases until a critical temperature is reached, triggering a phase transition. This causes the fluid to form a dense, crystalline lattice progressing from the center outward. This liquid-to-solid phase transition releases latent heat and temporarily slows cooling, which may be observable in the HR diagram as a bottleneck of white dwarfs \citep{vanhornCrystallizationWhiteDwarfs1968}. That said, the latent heat resulting from this process is small -- on the order of $kT$, where $T$ is the solidification temperature -- so the bottleneck is expected to be small as well \citep[as first noted by ][]{shavivCoolingCarbonoxygenWhite1976}.

Although latent heat release during crystallization has long been understood as a source of cooling delay, other mechanisms that may co-occur with crystallization can also contribute. For example, ``phase separation'' can take place within the WD interior -- this general class of process occurs when two distinct phases are created from a previously homogeneous mixture. This may result in the liberation of gravitational energy due to internal redistribution within the structure of the chemical density profiles. The phase separation of carbon and oxygen was initially suggested by \citet{stevensonEutecticCarbonOxygenWhite1980}, and applied to the cooling of white dwarfs by \citet{mochkovitchFreezingCarbonoxygenWhite1983}, with later additions by \citet{garcia-berroTheoreticalWhitedwarfLuminosity1988, hernanzInfluenceCrystallizationLuminosity1994, isernPhysicsCrystallizingWhite1997}.

Another distinct process that functions to liberate gravitational potential energy is ``gravitational settling''. This process was first applied to neutron-rich impurities in the WD core by \citet{bildstenGravitationalSettling22NE2001}. In this context, settling occurs when species such as $^{22} \mathrm{Ne}$ and $^{26} \mathrm{Mg}$ experience a net downward force as a result of their excess neutrons. These species may be present in WD interiors as a result of mergers between WDs and subgiant stars \citep{shenBranchCoolingAnomaly2023, tremblayGaiaWhiteDwarf2024}, or simply single-star evolution \citep{chengCoolingAnomalyHighmass2019}. Settling is a slow process that is effectively stopped by crystallisation due to the requirement for the neutron-rich isotopes to move freely through the liquid plasma. The combination of the phase transition and gravitational settling is expected to result in a cooling delay of $1-2$ Gyr, with $\sim 0.6$ Gyr contributed by C/O phase separation \citep[e.g.,][]{blouinPrecisionCosmochronologyNew2020, bauerCarbonOxygenPhase2023}.

With the release of the aforementioned Gaia DR2 and the identification of the Q branch, there was a rush of research activity to explain the observed properties of these peculiar WDs. \citet{chengCoolingAnomalyHighmass2019}, among others, proposed that the observed overdensity in this region of the HR diagram may be explained by the gravitational settling of $^{22} \mathrm{Ne}$. In particular, they noted that the total gravitational energy stored in $^{22} \mathrm{Ne}$ would in principle be sufficient to explain the cooling delay. However, this disagreed with existing numerical simulations, which found shorter delays for WDs with even very high $^{22} \mathrm{Ne}$ abundance \citep{camisassaEffect22NeDiffusion2016}. The lack of a clear explanation for the cooling delays observed in the Q branch presented a mystery. 

\citet{chengCoolingAnomalyHighmass2019} further investigated the population of Q branch WDs using the kinematic data provided by Gaia. This allowed for a disambiguation of the ages of the WDs contained within this population. Some ultra-massive WDs ($1.08-1.23$\msol) at the blue end of the Q branch, comprising $5 - 9$\% of the population, experience an extended cooling delay of $\geq 8$\,Gyr. 

These discoveries lead to a concerted research effort to identify the additional energy source powering the extended cooling delay in this stellar population. This included \citet{blouin22NePhaseSeparation2021} and \citet{blouinPhaseSeparationUltramassive2021}, where the authors modelled the full three-component CO/Ne phase diagram and suggested that a ``distillation'' mechanism \citep[an idea first proposed by][]{isernRoleMinorChemical1991} is at play. In short, \citet{blouin22NePhaseSeparation2021} employed Monte Carlo simulations to construct a three-component CO/Ne phase diagram. They predict that when crystals are formed, they will be depleted in $^{22} \mathrm{Ne}$ relative to the surrounding liquid. This causes the buoyant crystals to float outward while the denser $^{22} \mathrm{Ne}$-rich liquid sinks and eventually freezes at the centre. This ongoing redistribution significantly delays cooling by releasing extra gravitational energy.

Two regimes were identified: (i) for WDs with moderately high $^{22} \mathrm{Ne}$ abundances ($X \gtrsim 0.03$), as expected from some mergers or $\alpha$-enhanced progenitors, the outcome is a $^{22} \mathrm{Ne}$-rich central core and a multi-gigayear cooling delay \citep{barrientosFractionDistilledWhite2025}; (ii) for solar-like abundances ($X \sim 0.014$), the process forms a $^{22} \mathrm{Ne}$-rich shell instead, delaying cooling by roughly 1\,Gyr later in the WD's evolution. Binding energy calculations confirm that the energy released is sufficient to account for the observed Q branch excess, consistent with the $\sim 6$ per cent of ultramassive WDs inferred to undergo delayed cooling 
\citep{blouin22NePhaseSeparation2021}. \citet{bedardBuoyantCrystalsHalt2024} later confirmed this explicitly using full time-dependent element transport simulations using the \textsc{STELUM} code. The distillation explanation is also appealing because it can be shown to function efficiently over a wide range of metallicities \citep[see e.g.][]{salaris22NeDistillationCooling2024}. Thus, distillation can be invoked as an explanation for a wide variety of observed cooling delays.

\begin{figure*}
    \centering
    \includegraphics[width=\linewidth]{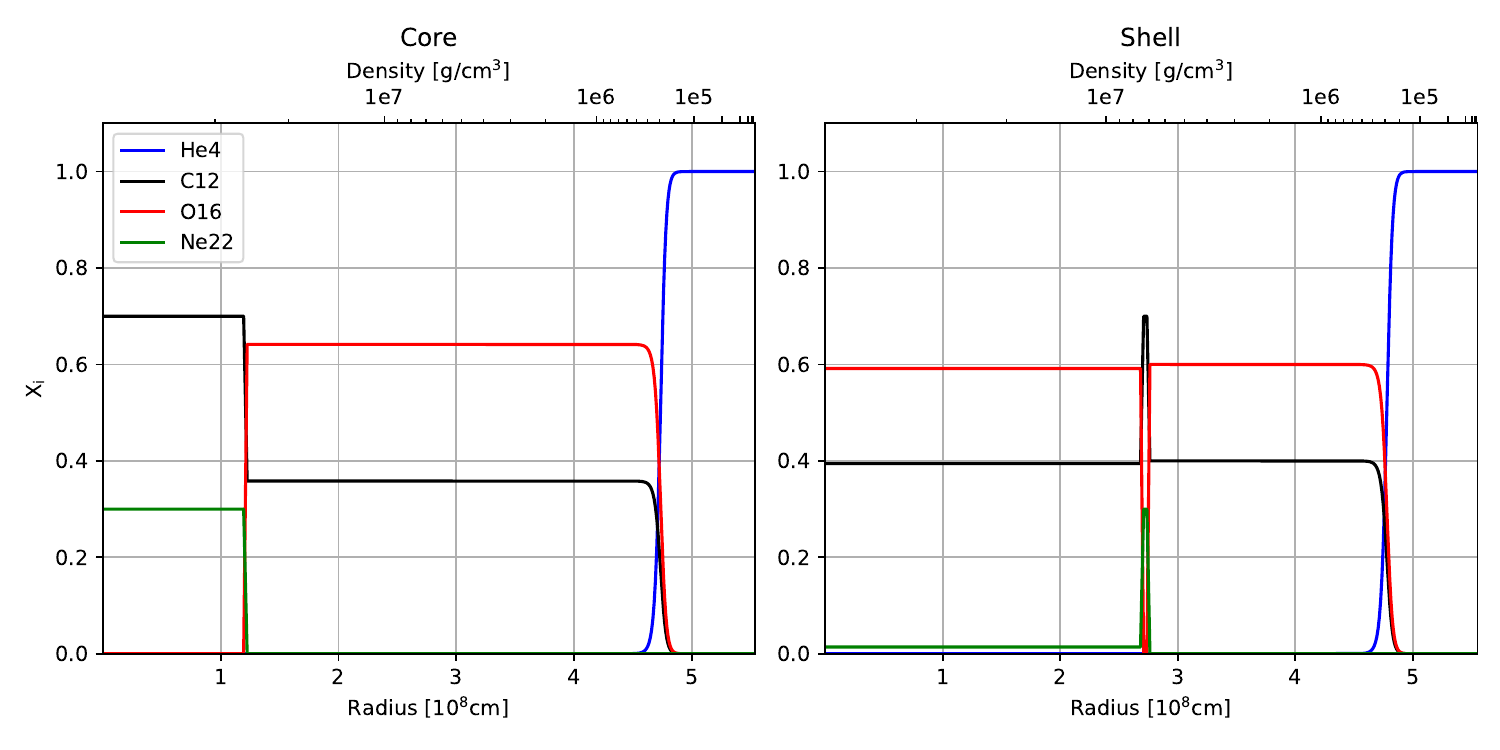}
    \caption{Comparison of the initial one-dimensional radial chemical profiles of the core and shell distilled models. Recall that the naming of these two profiles corresponds to the location of the majority of the $^{22} \mathrm{Ne}$ in the structure - i.e. the core distilled model, $M_c$, has a large volume of neon concentrated at the stellar core whereas the shell model, $M_s$, has a thin shell at a larger radius. Note that some density tick labels have been omitted in the atmosphere.}
    \label{fig:profiles}
\end{figure*}

\begin{table*}
\centering
\caption{The mass composition of the 3D WD models prior to detonation, but after the custom HEALPIX mapping, as discussed in Section~\ref{sec:methods}. The core and shell models are deonted by $M_{c}$ and $M_{s}$ respectively, while their homogeneous composition equivalents include the letter ``h''. All masses are given in solar masses. Note that while the total masses are different, the percentage differences in composition are low.}
\label{tab:init_composition}
\resizebox{0.90\linewidth}{!}{%
\begin{tabular}{lllllllll}
Species & $M_{c}$ & $M_{c}$ (\%) & $M_{hc}$ & $M_{hc}$ (\%) & $M_{s}$ & $M_{s}$ (\%) & $M_{hs}$ & $M_{hs}$ (\%) \\ \hline
He4 & 0.012 & 1.1 & 0.012 & 1.2 & 0.012 & 1.2 & 0.010 & 1.0 \\ 
C12 & 0.401 & 39.3 & 0.397 & 39.3 & 0.407 & 39.9 & 0.392 & 39.1 \\ 
O16 & 0.572 & 56.1 & 0.566 & 56.0 & 0.585 & 57.3 & 0.586 & 58.4 \\ 
Ne22 & 0.035 & 3.4 & 0.035 & 3.5 & 0.016 & 1.6 & 0.014 & 1.4 \\ 
\hline 
Total & 1.020 & - & 1.010 & - & 1.020 & - & 1.003 & -\\ 
\end{tabular}%
}
\end{table*}

\subsection{Explosion Models in Distilled White Dwarfs}

In this work, we consider the explosion of stars that contain $^{22} \mathrm{Ne}$-enriched regions formed via distillation during crystallisation. Our investigation follows the early studies of \citet{bravoContributionNe22Synthesis1992} that examined the role of $^{22} \mathrm{Ne}$ in contributing to the neutron excess in the central regions of the star, leading to the overproduction of neutron-rich isotopes. Later work explored whether variation in the $^{22} \mathrm{Ne}$ content of WD progenitors (arising from differences in progenitor metallicity) can affect the dynamical outcome of SNe\,Ia \citep{townsleyEvaluatingSystematicDependencies2009}, beyond the effect of increasing neutron excess which reduces $^{56} \mathrm{Ni}$ production \citep{timmesVariationsPeakLuminosity2003}. Such phase separation results in sharp chemical and electron-fraction gradients within the WD interior, which have the potential to significantly alter ignition conditions, detonation propagation, and nucleosynthesis yields. 

We consider two regimes that correspond to two different $^{22} \mathrm{Ne}$ abundances and, thus, lengths of cooling delay:

\begin{enumerate}
    \item moderately high abundance, leading to a $^{22} \mathrm{Ne}$-rich central core ($M_c$)
    \item solar-like abundance, leading to a $^{22} \mathrm{Ne}$-rich shell ($M_s$).
\end{enumerate}

To investigate these regimes, we carry out three-dimensional simulations with \textsc{AREPO} to determine how the geometry and distribution of $^{22}\mathrm{Ne}$ influence the thermonuclear runaway, particularly the resulting changes in nucleosynthetic yields and their effects on the emergent light curves and spectra. These models probe whether core enrichment leads to observationally distinct features in the light curves and spectra, and whether such events may correspond to rare or peculiar subclasses of SNe\,Ia.

In the shell-dominated regime, the middle layers of the crystallising CO WD become enriched with $^{22} \mathrm{Ne}$, forming a stratified structure atop a solid CO core. This shell introduces sharp gradients in both composition and density, which may influence detonation dynamics. It may act as a barrier to detonation propagation or alter shock structure and flame stability. Furthermore, the increased neutron excess in the shell favours the synthesis of stable iron-group elements such as $^{58} \mathrm{Ni}$ and $^{54} \mathrm{Fe}$ over $^{56} \mathrm{Ni}$, potentially modifying both nucleosynthetic yields and observable supernova signatures. In regions of high $^{22} \mathrm{Ne}$ abundance, flame speeds are reduced, which may further inhibit detonation propagation depending on the shell's thickness and local density.

In the core-dominated regime, which arises in WDs with moderately elevated $^{22} \mathrm{Ne}$ abundances, distillation leads to the accumulation of $^{22} \mathrm{Ne}$ in the central regions of the star. This results in a dense, neutron-rich core surrounded by a CO mantle. The high central neutron excess is expected to significantly impact the ignition process and nucleosynthesis, particularly by enhancing the production of neutron-rich iron-group isotopes in the innermost ejecta. The steep inward gradient in $Y_e$ may influence detonation convergence, potentially altering both the central burning conditions and flame geometry.

Some work exists already in the literature exploring the effects of chemical segregation on the characteristics of SNe\,Ia -- notably \citet{bravoTypeIaSupernovae2024} -- where the authors examine the effect of impurities such as $^{22} \mathrm{Ne}$ and $^{56} \mathrm{Fe}$. This overlaps with our research goals, however there are differences in our approaches. We examine different WD masses and different chemical profiles using an alternative coupled nuclear network. Additionally, we extend the analysis by computing the spectra of these detonations to give a greater focus on observables (see Section~\ref{sec:results}). 

To explore these effects, we have conducted 3D simulations using the \textsc{AREPO} moving-mesh code, initiating the detonation by imposing a localised temperature enhancement at the core. These simulations allow us to assess whether $^{22} \mathrm{Ne}$-rich stratification can lead to non-standard thermonuclear outcomes, including sub-luminous or peculiar SNe\,Ia, or whether the explosion proceeds in a manner consistent with standard SNe\,Ia.

\section{Methods} \label{sec:methods}

In order to investigate the behaviour of our distilled WD models, we first require chemical profiles to compare. These derive from the work of \citet{blouin22NePhaseSeparation2021}, and are shown in Figure~\ref{fig:profiles}. Both models have a total mass of approximately $1.0$\,\msol, but different $^{22} \mathrm{Ne}$ masses located in different regions within the stellar structure. 

The core model ($M_c$) contains a region of carbon and neutron-rich material -- i.e. $X(^{22}\mathrm{Ne}) = 0.30$; $X(^{12}\mathrm{C}) = 0.70$ -- completely within a radius of $r_{\text{core}} = 1.2 \times 10^8$\,\cm. The equivalent density at this radius is $\rho = 2.5 \times 10^7$\ \gcmcubed. In the middle layers, the star is purely carbon-oxygen. A helium layer is included with a mixed C/O/He region located at $r_{\text{atm}} \sim 4.5 \times 10^8$\cm\, ($\rho = 5.3 \times 10^5$\,\gcmcubed).

The shell model ($M_s$), has a neon mass fraction $X(^{22}\mathrm{Ne}) = 0.014$ up to an inner shell radius of $r_{\text{s,inner}} = 2.68 \times 10^8$\,\cm\, ($\rho = 7.5 \times 10^6$\,\gcmcubed). The shell itself contains 30\% $^{22}\mathrm{Ne}$ up to its outer radius, $r_{\text{s,outer}} = 2.76 \times 10^8$\,\cm\, ($\rho = 6.9 \times 10^6$\,\gcmcubed), at which point the $^{22} \mathrm{Ne}$ abundance drops to zero. This model also includes a helium layer with a mixed C/O/He region located at $r_{\text{atm}} \sim 4.5 \times 10^8$\cm\, ($\rho = 5.7 \times 10^5$\,\gcmcubed).

We note that recent works on the internal structure of core-distilled WDs have determined that a non-uniform $X(^{22}\mathrm{Ne})$ distribution is more likely \citep[see, e.g.,][]{salaris22NeDistillationCooling2024, bedardBuoyantCrystalsHalt2024}. However, the final composition profile remains uncertain, as discussed in the Methods section of \citet{bedardBuoyantCrystalsHalt2024}. Given the uncertainties involved, we consider that a uniform central composition is a reasonable proxy for the range of possible outcomes following the distillation process. Future studies may wish to explore how uniform versus non-uniform core compositions affect the resulting nucleosynthesis and dynamic.

For each distilled WD we have created a homogeneous equivalent that has the same overall composition, but with the neon distributed evenly at all radii, and up to a 100 per cent helium atmosphere. These models are denoted as $M_{hc}$ and $M_{hs}$ to indicate a homogeneous equivalent to the core model and shell model. To create the 1D profiles for the homogeneous equivalent models, we use an isothermal, constant composition approximation. Using a temperature of $T = 5 \times 10^5$\ \K, the structure is generated using a simple numerical integrator as discussed in more detail in \citet{pakmorStellarGADGETSmoothed2012}. The process requires estimating the central density of the star and then numerically integrating the hydrostatic equilibrium condition up to a cutoff density (in this case, $\rho_c = 1 \times 10^{-4}$\,\gcmcubed). The outer radii of these homogeneous models is set as 100 per cent helium to match the helium content of the distilled model. These four models will allow for a direct comparison that aims to isolate the effect of distillation, as well as a cross-comparison between the two distilled models of different metallicity. 

Beginning with our 1D chemical profiles, we required a technique to transform this into a structure compatible with the moving-mesh simulation code \textsc{AREPO}. We employed a modified HEALPIX (Hierarchical Equal Area isoLatitude Pixelisation) mapping \citep{gorskiHEALPixFrameworkHighResolution2005} to transform this 1D structure to 3D. This mapping treats every point in the one-dimensional profile as a radius and generates a series of points equally spaced in latitude and longitude. This allowed us to create a spherically symmetric set of points whose radial projections matched the initial profiles from \citet{blouin22NePhaseSeparation2021}. The specific modifications to the HEALPIX algorithm are described in \citet{pakmorStellarGADGETSmoothed2012} -- these involve allocating points in batches to ensure the 3D point cloud approximates cubes of constant density. The masses of the 3D WD structures for all four models is shown in Table~\ref{tab:init_composition}. Note that the mass fractions of carbon, oxygen, and helium have been chosen to be similar between the two distilled models. 

We make use of \textsc{AREPO}'s mass refinement capabilities for efficient computation. We activate explicit refinement and de-refinement when the mass of a cell is larger than twice or smaller than half of the target mass resolution. We have used a mass resolution of $5 \times 10^{-7}$\,\msol\, for these simulations in order to resolve the thin $^{22}\mathrm{Ne}$ shell. Additional refinement is triggered when the volume of a cell is more than 10 times larger than its smallest direct neighbour. This is necessary to avoid large resolution gradients in the mesh at steep density gradients. Our simulations include a ``background mesh'' of pure helium with a density of $10^{-4}$\,\gcmcubed \, to avoid numerical problems arising from steep gradients from the stellar surface to empty space. We also enforce a maximum cell volume of $10^{30} \mathrm{cm}^3$ to prevent de-refinement of the background mesh. We soften the gravitational force to avoid spurious two-body interactions with a softening length of 2.8 times the radius of a cell, but force the softening to be at least 10\,km. The calculation of the primitive variables (e.g. temperature, pressure) is provided by a non-ideal Helmholtz equation of state (HES) \citep{timmesAccuracyConsistencySpeed2000}. Moreover, we fully couple a 55-isotope nuclear reaction network \citep{pakmorThermonuclearExplosionMassive2021} -- this is the minimum network size available that includes the key $^{22}\mathrm{Ne}$ isotope. The activation criterion for the nuclear reaction network is described further in \citet{burmesterAREPOWhiteDwarf2023, seitenzahlSpontaneousInitiationDetonations2009}.

Our simulation consists of a two-step process: (1) Relaxation, and (2) Detonation. A short relaxation simulation is required due to the possibility of discretisation errors in the transformation from 1D to 3D which disrupt hydrostatic equilibrium. Thus, some relaxation method is required to ensure a stable stellar structure and prevent unphysical effects such as temperature spikes. We place the generated structure in the centre of a simulated box of size $10^{10}$\,\cm\, and damp away spurious velocities for the target WD in isolation for 10\,s (i.e. several dynamical timescales) using the criterion outlined in \citet{ohlmannConstructingStable3D2017}. As this simulation is ongoing, we check the stability of the stellar structure against the original 1D structures. In particular, we aim to preserve the location of the core and shell features as closely as possible. ``Passive Scalar'' labels are added to track the origin location of particles (e.g., within the shell, in the atmosphere). 

\begin{table*}
\centering
\caption{Asymptotic nucleosynthetic yields (in solar masses) of stable isotopes after a 2 Gyr decay.}
\label{tab:ejecta_stable}
\resizebox{\linewidth}{!}{
\begin{filecontents*}{data/table_ejecta_stable.txt}
Spec_0,MC_0,MHC_0,MS_0,MHS_0,Spec_1,MC_1,MHC_1,MS_1,MHS_1
C12,5.2826e-05,4.6199e-05,5.4090e-05,7.7992e-05,Ca46,1.2797e-13,4.3318e-10,3.2029e-14,1.7214e-11
C13,2.1154e-10,6.7795e-11,1.7170e-10,2.7820e-11,Ca48,1.3083e-16,9.8291e-15,1.2676e-18,3.8537e-17
N14,4.6270e-09,8.0809e-11,4.5046e-09,1.3924e-10,Sc45,2.2614e-07,2.6172e-06,3.1155e-07,2.7329e-06
N15,1.2609e-06,1.2380e-11,1.2991e-06,8.8960e-11,Ti46,2.2751e-07,2.6386e-05,2.6326e-06,1.0055e-05
O16,7.5860e-02,8.0262e-02,7.5349e-02,8.7970e-02,Ti47,9.7094e-06,3.4014e-05,9.6022e-06,2.5105e-05
O17,1.9378e-12,8.1863e-12,2.7966e-12,2.3286e-11,Ti48,1.5453e-03,1.4614e-03,1.4224e-03,1.2671e-03
O18,5.8480e-14,2.3359e-12,5.5664e-14,5.9776e-12,Ti49,3.7124e-06,4.0428e-05,2.3655e-05,3.1065e-05
F19,6.3025e-13,1.2182e-14,6.3580e-13,3.3652e-14,Ti50,1.2806e-11,2.2518e-09,1.7947e-11,3.5520e-10
Ne20,7.0569e-05,3.2812e-05,7.2750e-05,1.1400e-04,V50,6.7526e-12,5.8860e-09,2.9350e-10,2.5947e-09
Ne21,3.7916e-10,9.3374e-10,4.9041e-10,1.5974e-09,V51,3.7314e-05,1.5292e-04,9.0769e-05,9.3861e-05
Ne22,9.5576e-10,1.7195e-09,1.2010e-09,2.4259e-09,Cr50,4.5491e-06,4.7070e-04,3.3177e-04,1.6959e-04
Na23,1.2011e-06,1.1463e-06,1.2687e-06,1.4620e-06,Cr52,1.1766e-02,8.7238e-03,1.0184e-02,9.6560e-03
Mg24,9.0188e-03,2.4346e-03,9.6225e-03,4.4526e-03,Cr53,1.6827e-04,9.2217e-04,6.5194e-04,6.9209e-04
Mg25,4.9756e-08,2.3938e-06,5.6085e-08,3.7491e-06,Cr54,9.1819e-10,1.4416e-07,6.6145e-08,1.4700e-08
Mg26,9.6792e-08,3.6400e-06,1.1334e-07,3.3692e-06,Mn55,1.5207e-03,5.4876e-03,3.3168e-03,3.5375e-03
Al27,2.0419e-05,1.6400e-04,2.1558e-05,1.7490e-04,Fe54,1.6113e-02,4.3326e-02,2.5788e-02,1.7810e-02
Si28,2.0830e-01,2.3024e-01,2.2150e-01,2.3602e-01,Fe56,3.9784e-01,4.1270e-01,4.4656e-01,4.1109e-01
Si29,4.0218e-05,4.8740e-04,4.1532e-05,3.2032e-04,Fe57,6.4386e-03,9.8232e-03,6.9506e-03,5.9296e-03
Si30,3.1682e-05,1.4639e-03,2.2440e-05,5.4412e-04,Fe58,2.0917e-09,4.9065e-08,1.7744e-08,6.0956e-09
P31,5.8522e-05,4.9329e-04,5.0128e-05,2.9491e-04,Co59,3.2780e-04,2.7460e-04,1.9794e-04,1.3512e-04
S32,1.4633e-01,1.2975e-01,1.4059e-01,1.4893e-01,Ni58,7.1683e-02,2.0383e-02,7.8288e-03,6.2374e-03
S33,3.1215e-05,3.6823e-04,4.2864e-05,2.5619e-04,Ni60,1.8916e-03,2.8248e-03,4.0729e-03,2.5900e-03
S34,3.1708e-06,3.9119e-03,2.8600e-05,1.4945e-03,Ni61,3.5375e-05,1.3286e-04,1.3126e-04,8.1168e-05
S36,1.0907e-10,1.2260e-06,9.1597e-10,4.6106e-07,Ni62,2.3053e-04,1.4796e-03,7.7835e-04,4.9859e-04
Cl35,9.1286e-06,1.3397e-04,1.0315e-05,8.5080e-05,Ni64,1.2303e-14,5.2091e-14,2.1296e-14,8.7803e-15
Cl37,2.9722e-06,3.6104e-05,5.4503e-06,2.5802e-05,Cu63,2.3480e-06,1.2420e-06,1.6380e-06,2.4563e-07
Ar36,3.1462e-02,2.2175e-02,2.8283e-02,2.8201e-02,Cu65,4.1775e-07,1.0489e-06,1.2334e-06,5.9121e-07
Ar38,5.8523e-07,1.9208e-03,2.3262e-05,7.1868e-04,Zn64,2.5104e-05,7.1252e-06,1.4526e-05,6.1823e-06
Ar40,2.6606e-11,4.5426e-08,9.7242e-11,2.0186e-08,Zn66,2.9871e-06,1.9963e-05,1.3314e-05,7.2736e-06
K39,6.9550e-06,1.3912e-04,1.2061e-05,8.8594e-05,Zn67,7.3582e-09,2.6421e-08,2.0845e-08,4.8838e-09
K41,1.1563e-06,9.0521e-06,1.6680e-06,9.0315e-06,Zn68,3.7444e-09,1.3256e-08,6.1352e-09,1.5912e-09
Ca40,3.1930e-02,1.9895e-02,2.7590e-02,2.6368e-02,Zn70,1.8745e-17,1.2647e-16,1.1432e-16,3.9207e-18
Ca42,2.1272e-07,6.3338e-05,7.3112e-07,2.3804e-05,Ga69,1.5899e-11,1.0442e-10,1.5696e-10,1.9948e-12
Ca43,1.9592e-06,3.5262e-06,2.1984e-06,3.0643e-06,Ga71,1.3868e-12,1.1560e-11,1.6889e-11,1.5132e-13
Ca44,4.8472e-04,5.4766e-04,4.2859e-04,5.9523e-04,-,-,-,-,-
\end{filecontents*}
\pgfplotstabletypeset[
multicolumn names,
col sep=comma,
columns=
{Spec_0,MC_0,MHC_0,MS_0,MHS_0,Spec_1,MC_1,MHC_1,MS_1,MHS_1},
columns/Spec_0/.style={string type, column name={Species},},
columns/MC_0/.style={string type, column name={$M_{c}$},},
columns/MHC_0/.style={string type, column name={$M_{hc}$},},
columns/MS_0/.style={string type, column name={$M_{s}$},},
columns/MHS_0/.style={string type, column name={$M_{hs}$},column type={c|}},
columns/Spec_1/.style={string type, column name={Species},},
columns/MC_1/.style={string type, column name={$M_{c}$},},
columns/MHC_1/.style={string type, column name={$M_{hc}$},},
columns/MS_1/.style={string type, column name={$M_{s}$},},
columns/MHS_1/.style={string type, column name={$M_{hs}$},column type={c}},
every head row/.style={before row=\toprule, after row=\midrule},
every last row/.style={after row=\bottomrule},
]{data/table_ejecta_stable.txt}
}
\end{table*}

In the detonation step, we insert the relaxed stellar structure at the centre of a larger simulated box of size $10^{12}$\,\cm. At this stage, we add one million Lagrangian tracer particles, which are tracked by the \textsc{AREPO} code as a separate particle type. These tracer particles are advected with the flow and record their temperature, pressure, and density as the simulation progresses. These will be used for post-processing to obtain the detonation yield according to the 384-species nuclear network, as described in Section~\ref{sec:384_spec_network}. Lastly, we locate the particle closest to the centre of the simulation box and raise its temperature to $3\,\mathrm{GK}$ to trigger a detonation. The burning front proceeds outwards supersonically ($M \sim 2.2$), consuming the available fuel within $\sim 1$\,\s. We allow the simulation to progress for $50.0$\ \s, until the resulting ejecta are in a state of homologous expansion.

\section{Results} \label{sec:results}

\subsection{Relaxation phase} \label{sec:results_relaxation}

The relaxation phase yielded a stable 3D configuration whose radial structure closely follows the 1D phase-separation profiles of \citet{blouin22NePhaseSeparation2021}. Both the $M_c$ and $M_s$ models maintained hydrostatic equilibrium, and we observed minimal spreading of the sharp shell feature in $M_s$. As is to be expected, there was some refinement of the WD structure around the atmosphere, softening the density gradient from star to background. This is due to the fact that there are only a handful of cells available to represent the low-density region. The 3D mass and percentage figures in Table~\ref{tab:init_composition} match the 1D profiles very closely (i.e., to $< 0.5\%$). The mass fractions of the distilled models also closely match those of their homogeneous equivalents. Thus, the relevant aspects of the structure are well-reproduced by our \textsc{AREPO} models.

\begin{table*}
\centering
\caption{Nucleosynthetic yields (in solar masses) of select radioactive isotopes at time $t = 50$\,\s.}
\label{tab:ejecta_unstable}
\resizebox{\linewidth}{!}{
\begin{filecontents*}{data/table_ejecta_unstable.txt}
Spec_0,MC_0,MHC_0,MS_0,MHS_0,Spec_1,MC_1,MHC_1,MS_1,MHS_1
C14,3.4964e-16,3.7073e-12,3.1132e-16,8.1140e-12,Mn53,3.8408e-06,5.0974e-05,4.3721e-05,1.8794e-05
Na22,9.0674e-10,7.6356e-10,1.1277e-09,9.8784e-10,Mn54,8.8046e-10,1.4381e-07,6.6073e-08,1.4614e-08
Al26,9.3108e-08,2.1043e-07,1.0801e-07,7.3514e-07,Fe52,1.1762e-02,8.6664e-03,1.0143e-02,9.6479e-03
Si32,5.8434e-14,3.5428e-10,2.0004e-13,3.2090e-11,Fe53,1.6442e-04,8.7119e-04,6.0821e-04,6.7329e-04
P32,2.0061e-11,4.2070e-07,1.2667e-10,1.2201e-07,Fe55,2.9655e-05,1.6248e-04,1.4699e-04,2.5168e-05
P33,2.4490e-11,3.9977e-07,6.3207e-11,1.0536e-07,Fe59,1.9392e-15,7.8313e-16,1.0577e-16,3.4629e-16
S35,1.3466e-11,1.9358e-07,4.5838e-11,6.1828e-08,Fe60,5.5246e-18,7.0955e-17,2.2838e-18,1.6417e-17
Cl36,9.5333e-11,1.0121e-06,8.8641e-10,4.3318e-07,Co55,1.4911e-03,5.3250e-03,3.1698e-03,3.5123e-03
Ar37,2.9721e-06,3.4982e-05,5.4499e-06,2.5427e-05,Co56,3.9719e-05,2.4718e-05,1.7728e-05,1.2940e-05
Ar39,8.7096e-13,7.8037e-09,4.3628e-12,2.9831e-09,Co57,2.5501e-05,2.1939e-05,1.9637e-05,3.8461e-06
K40,2.6199e-11,3.9004e-08,9.6459e-11,1.9535e-08,Co58,2.0272e-09,4.6978e-08,1.7441e-08,5.9346e-09
Ca41,1.1563e-06,9.0457e-06,1.6680e-06,9.0295e-06,Co60,4.7405e-15,3.5328e-13,2.5185e-14,4.0565e-14
Ti44,4.8472e-04,5.4743e-04,4.2859e-04,5.9516e-04,Ni56,3.9779e-01,4.1227e-01,4.4632e-01,4.1103e-01
V48,1.5823e-07,5.2912e-07,1.7887e-07,3.4663e-07,Ni57,6.4131e-03,9.8012e-03,6.9310e-03,5.9258e-03
V49,1.7632e-08,4.4747e-07,1.3868e-07,2.2358e-07,Ni59,1.3488e-04,3.8849e-05,2.7183e-05,1.8019e-05
Cr48,1.5451e-03,1.4605e-03,1.4223e-03,1.2666e-03,Ni63,4.8026e-16,4.8863e-15,3.1681e-16,5.4777e-16
Cr49,3.6947e-06,3.9978e-05,2.3516e-05,3.0841e-05,Zn62,2.3051e-04,1.4794e-03,7.7826e-04,4.9853e-04
Cr51,1.4388e-07,3.9681e-06,2.9827e-06,1.0123e-06,Zn65,1.4078e-09,3.8949e-09,4.3893e-09,2.3289e-09
Mn51,3.7170e-05,1.4893e-04,8.7785e-05,9.2843e-05,Ge65,1.7801e-07,4.2575e-07,5.1148e-07,2.3126e-07
Mn52,3.1203e-06,4.6229e-06,4.2517e-06,3.5636e-06,Ge68,3.7444e-09,1.3255e-08,6.1351e-09,1.5912e-09
\end{filecontents*}
\pgfplotstabletypeset[
multicolumn names,
col sep=comma,
columns=
{Spec_0,MC_0,MHC_0,MS_0,MHS_0,Spec_1,MC_1,MHC_1,MS_1,MHS_1},
columns/Spec_0/.style={string type, column name={Species},},
columns/MC_0/.style={string type, column name={$M_{c}$},},
columns/MHC_0/.style={string type, column name={$M_{hc}$},},
columns/MS_0/.style={string type, column name={$M_{s}$},},
columns/MHS_0/.style={string type, column name={$M_{hs}$},column type={c|}},
columns/Spec_1/.style={string type, column name={Species},},
columns/MC_1/.style={string type, column name={$M_{c}$},},
columns/MHC_1/.style={string type, column name={$M_{hc}$},},
columns/MS_1/.style={string type, column name={$M_{s}$},},
columns/MHS_1/.style={string type, column name={$M_{hs}$},column type={c}},
every head row/.style={before row=\toprule, after row=\midrule},
every last row/.style={after row=\bottomrule},
]{data/table_ejecta_unstable.txt}
}
\end{table*}

The central densities of both the core and shell 3D structures increase somewhat as a result of relaxation. The core model has an initial central density of $\rho = 3.81 \times 10^7$\ \gcmcubed\ which increases to $\rho = 4.17 \times 10^7$\ \gcmcubed\ ($+ 9.4 \%$). Likewise the shell model has an initial central density of $\rho = 3.47 \times 10^7$\ \gcmcubed\ which increases to $\rho = 3.82 \times 10^7$\ \gcmcubed\ ($+ 10 \%$). The homogeneous equivalent models experienced a less significant increase in central density ($< 1\%$).

\subsection{Detonation phase (55-species network)} \label{sec:results_detonation}

Our triggering procedure was successful in igniting a runaway thermonuclear explosion in the core. The burning region quickly encompasses the entire WD, with the majority of the burning completed by $t = 0.5$\ \s\, and the ejecta expanding freely by $t = 1$\ \s. This continues through to the end of the simulation at $t = 50$\ \s.

\subsection{384-isotope nuclear reaction network} \label{sec:384_spec_network}

The yields captured by the 55-isotope network can be refined by post-processing with the more complete 384-species reaction network. This is done using the data collected by the one million Lagrangian tracers, which are advected with the flow during the detonation phase and through to homologous expansion. These particles continuously record their density, temperature, and pressure, which can be used to replay their evolutionary trajectory using the additional nuclear transitions captured by the larger network size \citep[for further detail see, e.g.,][]{burmesterAREPOWhiteDwarf2023}. We retain the 55-species network yields, but have not used them in subsequent sections unless stated explicitly. This post-processing is made possible by the ``Yet Another Nuclear Network'' (\textsc{YANN}) code \citep{pakmorStellarGADGETSmoothed2012}. We also consider the yield of the explosion after a period of radioactive decay --- in this case, 2 Gyr. To determine the late time isotopic abundances, we assume that all radioactive isotopes with a half-life shorter than 2 Gyr have completely decayed to their stable products.

We present the results from the network post-processing in Tables~\ref{tab:ejecta_stable} and \ref{tab:ejecta_unstable}. Table~\ref{tab:ejecta_unstable} gives the mass values of some abundant and long-lived radioactive isotopes at $t = 50$\,\s\,, whereas Table~\ref{tab:ejecta_stable} gives the yields of stable isotopes. Those isotopes with half-lives longer than 2 Gyr are listed with their production yields as at $t = 50$\,\s.

\subsubsection{Comparing core distilled to shell distilled}

The $^{56} \mathrm{Ni}$ in all cases is consistent with moderately luminous SNe\,Ia: $0.398$\,\msol\, (core) and $0.446$\,\msol\,(shell). We also note the dominance of $\alpha$-chain elements such as $^{28} \mathrm{Si}$ ($0.208$\,\msol, $M_c$; $0.222$\,\msol, $M_s$), $^{32} \mathrm{S}$ ($0.146$\,\msol; $0.141$\,\msol), $^{36} \mathrm{Ar}$, and $^{40} \mathrm{Ca}$, reflecting efficient silicon burning. There are comparable levels of intermediate-mass elements (IMEs) in both models, with slightly enhanced Si and Mn in the $M_s$ model.

Comparing the $M_c$ and $M_s$ models, we notice several key differences:

\begin{enumerate}
    \item $^{58} \mathrm{Ni}$ is significantly enhanced in $M_c$ ($0.0717$\,\msol) relative to $M_s$ ($0.00783$\,\msol), reflecting the higher central neutron excess in $M_c$.
    \item $^{55} \mathrm{Co}$, modestly present ($0.00149$\,\msol) in $M_c$, is more abundant in $M_s$ ($0.00317$\,\msol), consistent with incomplete silicon burning in neutron-rich middle layers.
    \item Other neutron-rich isotopes such as $^{54} \mathrm{Fe}$ ($0.0161$ \msol\, in $M_c$ vs. $0.0258$ \, \msol\, in $M_s$), $^{59} \mathrm{Ni}$, and $^{57} \mathrm{Fe}$ indicate that $M_c$ favours enhanced stable iron-group production.
\end{enumerate}

These trends reflect differences in neutron excess distribution and ignition geometry. In $M_c$, neutronisation is concentrated at high densities near the core, permitting complete processing. In $M_s$, the neutron-rich shell undergoes partial burning, contributing to the production of neutron-rich isotopes like $^{55} \mathrm{Co}$ and $^{54} \mathrm{Fe}$ in the outer ejecta. 

\begin{figure*}
    \centering
    \includegraphics[width=0.95\linewidth]{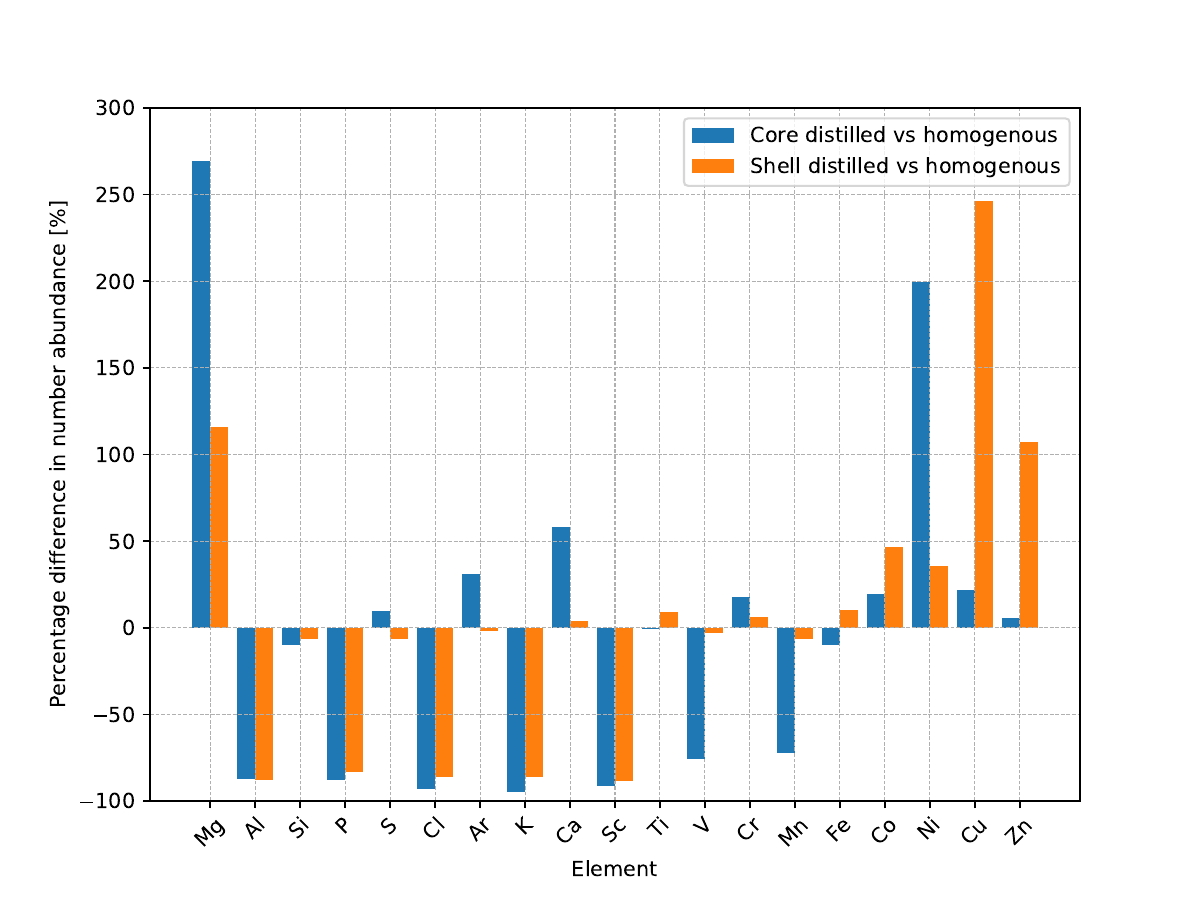}
    \caption{Number abundance for selected elements of distilled vs homogeneous models 2 Gyr after detonation. For example, a value of 100 per cent indicates that the distilled model produced twice as many atoms of a given element by number.}
    \label{fig:bar_number_abundance}
\end{figure*}

\subsubsection{Comparing distilled to homogeneous}

We compare our distilled models to standard sub-Chandrasekhar CO WD detonations with homogeneous compositions and no prior phase separation. For a progenitor mass of $\sim 1.0$\,\msol, such models typically produce $\sim 0.3 - 0.5$\,\msol\ of $^{56} \mathrm{Ni}$ \citep[e.g.,][]{simDetonationsSubChandrasekharmassC+O2010, shenSubChandrasekharmassWhiteDwarf2018}, consistent with the $^{56} \mathrm{Ni}$ yields recovered from all models. However, the distribution of neutron-rich isotopes differs markedly. 
In standard homogeneous models, the synthesis of $^{58} \mathrm{Ni}$ and $^{54} \mathrm{Fe}$ is limited unless the progenitor has super-solar metallicity, which increases the initial neutron excess via elevated $^{22} \mathrm{Ne}$ \citep{seitenzahlThreedimensionalDelayeddetonationModels2013, leungExplosiveNucleosynthesisSubChandrasekharmass2020}. By contrast, in our $M_c$ model, the central $^{22} \mathrm{Ne}$ enhancement leads to strong local neutronisation and efficient synthesis of stable neutron-rich isotopes, including $\sim 0.072$\,\msol\, of $^{58} \mathrm{Ni}$ and elevated $^{54} \mathrm{Fe}$. In the $M_s$ model, the neutron-rich shell undergoes incomplete burning, enabling the survival of $^{55} \mathrm{Co}$ at a mass of $3.2 \times 10^{-3}$\,\msol. While this is less than the homogeneous equivalent ($3.5 \times 10^{-3}$\,\msol), it is notable that the neutron-rich shell is located at a density of $\sim 7.4 \times 10^6$\,\gcmcubed. This is the density range at which we might expect a large mass of $^{55} \mathrm{Co}$ created via incomplete silicon burning to survive through to freeze-out. Experimenting with modifications to this shell distilled model (e.g., higher $^{22} \mathrm{Ne}$ mass fraction in the shell, a thicker shell) could prove interesting with respect to the production of $^{55} \mathrm{Mn}$ via decay of this isotope.

\begin{figure*}
    \centering
    \includegraphics[width=\linewidth]{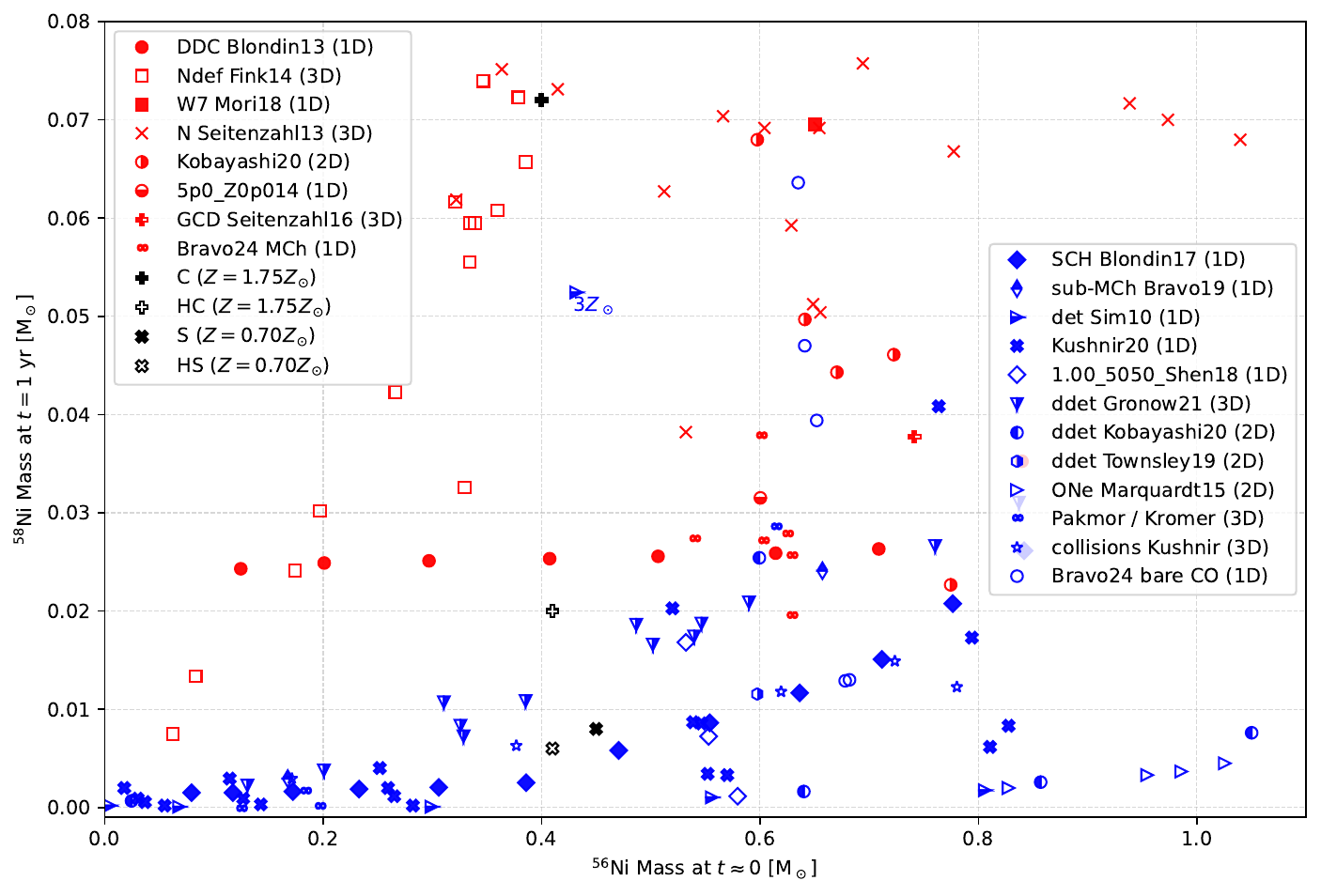}
    \caption{Figure partially reproduced from \citet{blondinStableNickelProduction2022} - Stable $^{58} \mathrm{Ni}$ yields one pear post explosion versus the radioactive $^{56} \mathrm{Ni}$ yield at $t \approx 0$ for various SN\,Ia models. The $M_{Ch}$ models are shown in red, while the sub-$M_{Ch}$ models are shown in blue. These include deflagrations, double detonations, and gravitationally confined detonations. Note also that simulations have been divided into 1D, 2D, and 3D. Our new results are shown in black.}
    \label{fig:blondin_comparison}
\end{figure*}

\begin{figure}
    \centering
    \includegraphics[width=\linewidth]{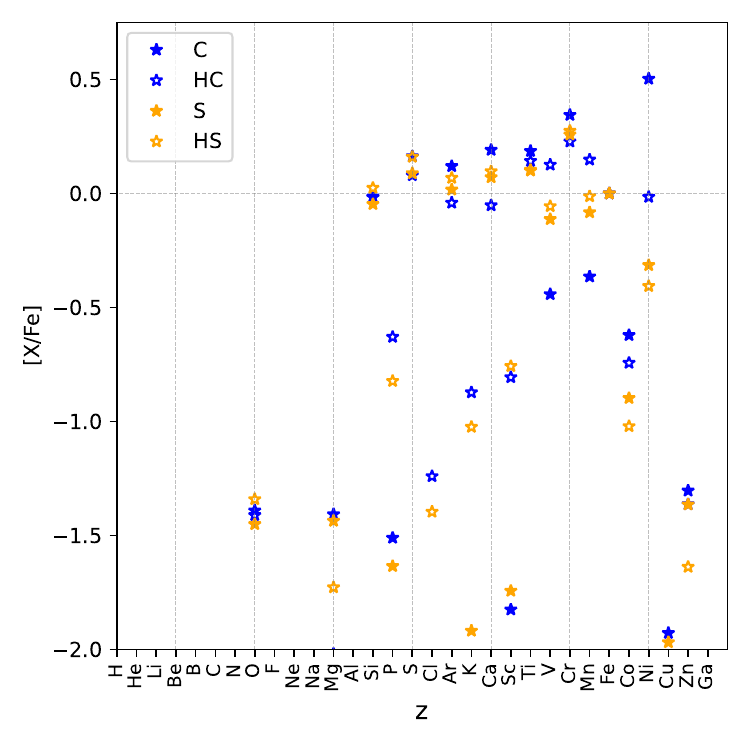}
    \caption{Comparison of elemental composition of the ejecta relative to iron and solar composition 2 Gyr after detonation, i.e. taking into account radioactive decay of unstable isotopes in the ejecta. We use the solar elemental abundances of \citet{asplundChemicalCompositionSun2009}.}
    \label{fig:abundance_ratio}
\end{figure}

These distinctions, driven solely by the redistribution of $^{22} \mathrm{Ne}$, highlight the potential for phase-separated progenitors to mimic some features of high-metallicity or near-Chandrasekhar-mass explosions, while remaining distinct from a dynamical and nucleosynthetic viewpoint. Using the mass data presented in Tables~\ref{tab:ejecta_stable} and \ref{tab:ejecta_unstable}, we are able to convert these mass values to number abundances. We have shown the percentage difference in number abundance in Figure~\ref{fig:bar_number_abundance} for selected elements. Several species show large differences in number abundance: notably nickel, which has a much higher $^{58} \mathrm{Ni}$ abundance compared to the homogeneous case. Calcium, too, is much more abundant in the core distilled case, driven by a significant increase in the production of $^{40} \mathrm{Ca}$. This is likewise the case for Argon, which is dominated by an increase in $^{36} \mathrm{Ar}$ despite lower masses in other isotopes. 

Not pictured in Figure~\ref{fig:bar_number_abundance} is nitrogen, which shows a very large enhancement in $^{15} \mathrm{N}$ in the shell distilled model ($1.3 \times 10^{-6}$\, versus $9.0 \times 10^{-11}$\,\msol). We find the $^{15} \mathrm{N} / ^{14} \mathrm{N}$ ratios for all models to be 10.7 (core), 0.01 (homogeneous core), 12.0 (shell), and 0.04 (homogeneous shell). This result may be interesting from an observational perspective with respect to dust and planetary formation, but we expect that it would be challenging to observe. If \ion{N}{I} is the dominant ion, signatures would be found primarily at $\lambda < 1800$\,\AA, which would be challenging to detect. We have determined that the nitrogen enhancement observed in the distilled models results from the decay of $^{15} \mathrm{O}$. Production of this proton-rich element is suppressed in the homogeneous WD models due to the presence of excess neutrons at all radii. We confirmed this by simulating the yields of a $1.0$\,\msol\, WD with no $^{22} \mathrm{Ne}$ content - this produced an equivalent $^{15} \mathrm{O}$ yield to the distilled models. 

\subsection{Comparison with other simulated yields} \label{sec:yield_comparison}

We now compare the yields of our simulations to various other sources in the literature. Figure~\ref{fig:blondin_comparison} is partially reproduced from \citet{blondinStableNickelProduction2022} and shows the comparison of stable $^{58} \mathrm{Ni}$ production to the radioactive $^{56} \mathrm{Ni}$ production from various previous simulations. The prior simulations include Chandrasekhar-mass and sub-Chandrasekhar-mass models of SNe\,Ia under various progenitor scenarios, including double detonations, deflagrations, and gravitationally confined detonations. The number of spatial dimensions is also indicated by the figure legend. One main conclusion of \citet{blondinStableNickelProduction2022} is the efficient separation of the $M_{ch}$ models (red) from the sub-$M_{ch}$ models (blue) by the combination of the radioactive and stable nickel yields. That is, low radioactive nickel and high stable nickel are a telltale sign of a $M_{ch}$ model, and vice versa. One exception to this rule is the ``Sim10 (1D)'' result, which boasts a high $^{58} \mathrm{Ni}$ mass but only moderate $^{56} \mathrm{Ni}$. However, this case is considered extreme due to the very high progenitor metallicity of $3 Z_{\odot}$. 

However, the inclusion of the core distilled model complicates this picture. It appears at a significantly higher stable nickel mass with an even lower progenitor metallicity of $1.75 Z_{\odot}$. This may indicate that the ansatz of using stable nickel as a discriminator between sub-$M_{ch}$ and $M_{ch}$ will fail when applied to WDs which have undergone significant distillation. Note that we have concluded the progenitor metallicity is well approximated by the mass fraction of $^{22} \mathrm{Ne}$ of the distilled WD. This has been validated by previous experiments with stellar evolution codes \citep[see, e.g.,][and references therein]{salarisUpdatedBASTIStellar2022}. Note also that have used a solar metallicity value of $0.02$ \citep{vagnozziSolarModelsLight2017} to compute the metallicity with respect to solar. Unlike the core model, the shell, homogeneous shell, and homogeneous core models are all consistent with the other sub-$M_{ch}$ models included for comparison.

\begin{figure}
    \centering
    \includegraphics[width=\linewidth]{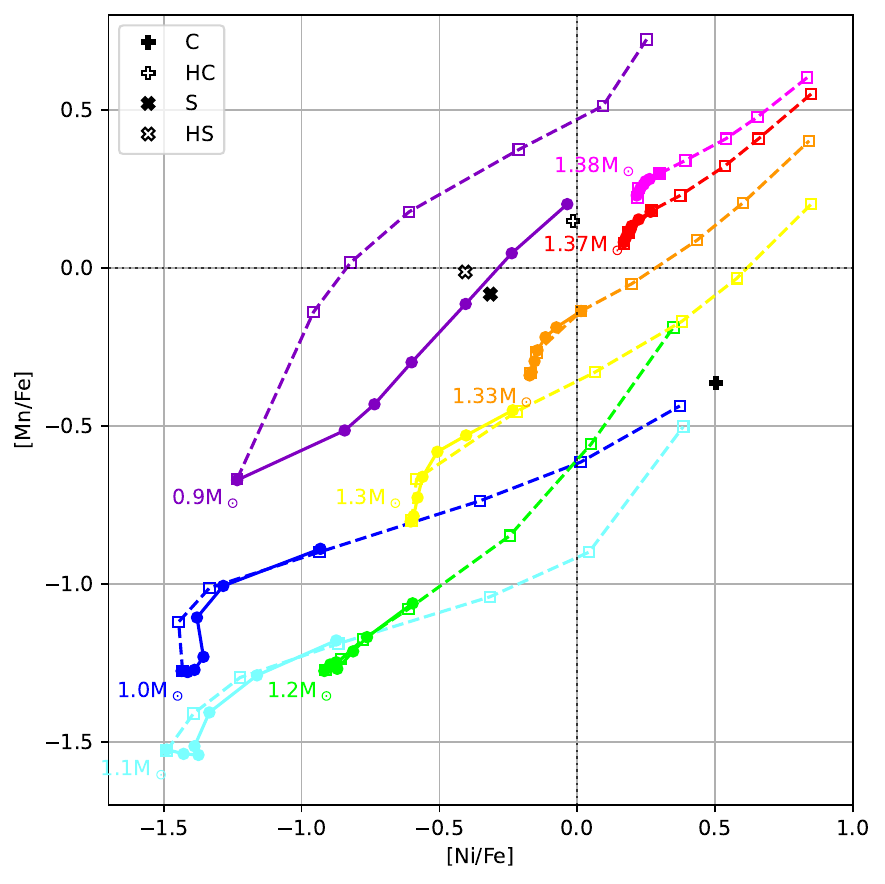}
    \caption{Figure partially reproduced from \citet{kobayashiNewTypeIa2020} - the Ni-Mn diagram constraining SN\,Ia enrichment. Results are shown for near-$M_{Ch}$ SNe with masses $M_{WD} = 1.38$, $1.37$, $1.33$, and $1.30$. The points indicate initial metallicities of $Z = 0$, $0.002$, $0.01$, $0.02$, $0.04$, $0.06$, and $0.10$ (left to right). Sub-$M_{Ch}$ SNe\,Ia are also given with masses $M_{WD} = 0.9$, $0.1$, $1.1$, and $1.2$ and metallicities $Z = 0$, $0.001$, $0.002$, $0.004$, $0.01$, $0.02$, and $0.04$ (left to right). The solid lines with filled circles are solar-scaled initial compositions originating from \citet{kobayashiNewTypeIa2020}, whereas the dashed lines with open squares are $^{22}\mathrm{Ne}$ only yields from \citet{leungExplosiveNucleosynthesisNearChandrasekharmass2018,leungExplosiveNucleosynthesisSubChandrasekharmass2020}. Our new results are shown in black.}
    \label{fig:kobayashi_comparison}
\end{figure}

Figure~\ref{fig:abundance_ratio} contains a comparison of the yield of our simulations at 2 Gyr in relation to solar levels using the solar abundance data from \citet{asplundChemicalCompositionSun2009}. Our neutron-rich WDs push production of many elements above solar levels, with most of the interesting elements among the transition metals (e.g., Si, S, Ar, and Ca). We are most interested in elements that equal or exceed solar abundance, and we must also pay special attention to cases where the distilled model exceeds the production of its homogeneous counterpart. The shell distilled model appears to produce less of any given element than its homogeneous counterpart, excluding some very low-production elements such as Mg, Cu, and Zn. One exception is the nickel production, where the shell distilled model outperforms its counterpart in the production of all stable nickel isotopes (cf. Table~\ref{tab:ejecta_stable}). The core distilled model is also superior in terms of its nickel abundance ratio, with a value of 0.50 compared to -0.02 in the homogeneous case. The core distilled model also shows strong enhancement of Ca and Ar, consistent with the increased number abundance seen in Figure~\ref{fig:bar_number_abundance}.

Figure~\ref{fig:kobayashi_comparison} has been partially reproduced from \citet{kobayashiNewTypeIa2020}. This figure shows the Ni-Mn diagram constraining SN\,Ia enrichment for sub-$M_{Ch}$ and near-$M_{Ch}$ models taken from \citet{kobayashiNewTypeIa2020} and \citet{leungExplosiveNucleosynthesisNearChandrasekharmass2018, leungExplosiveNucleosynthesisSubChandrasekharmass2020}, with the points indicating various metallicities. The $M_s$ and $M_{hs}$ models show only moderate differences (i.e., slight increases in nickel production coupled with slight decreases in manganese production), making them most analogous to a $0.9$\,\msol\, WD (purple) with a metallicity between $Z = 0.01 - 0.02$. The core distilled model shows significantly more pronounced differences when compared to the homogeneous counterpart. Whereas $M_{hc}$ has abundance ratios similar to a $0.9$\,\msol WD with a high metallicity ($Z = 0.03 - 0.04$), the core distilled model has abundance ratios similar to a $1.0 - 1.1$\,\msol\, model with $Z \gtrsim 0.04$.

\begin{figure*}
    \centering
    \includegraphics[width=0.85\linewidth]{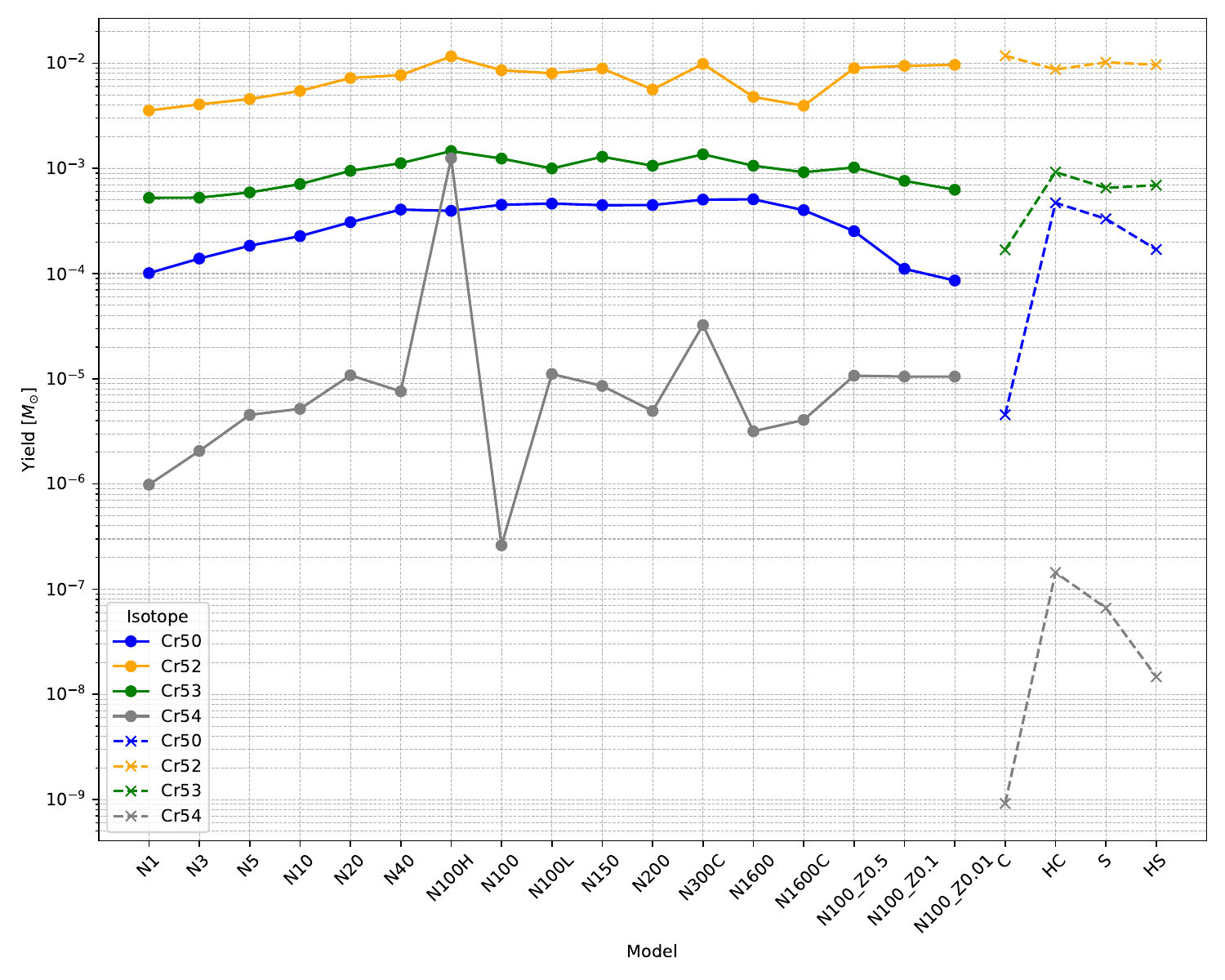}
    \caption{Chromium isotopic yields of our four WD models (right-hand side) compared to the Chandrasekhar-mass isotopic yields from \citet{seitenzahlThreedimensionalDelayeddetonationModels2013}. The various comparison models correspond to simulations with different numbers of ignition kernels, as well as a high and low central density.}
    \label{fig:chromium}
\end{figure*}

Chromium is another element that shows intriguing properties when compared with previous results. Figure~\ref{fig:chromium} shows the yields of our four models compared to the yields recorded in the Chandrasekhar-mass simulations in \citet{seitenzahlThreedimensionalDelayeddetonationModels2013} with respect to several isotopes of chromium. For all models, we observe high $^{52} \mathrm{Cr}$ production, equalling or exceeding the high metallicity models as well as those with a large number of ignition sites (indicated by the number after the ``N'' in the model name). Whereas the homogeneous core simulation yields high abundances of $^{50} \mathrm{Cr}$ and $^{53} \mathrm{Cr}$ and relatively less $^{52} \mathrm{Cr}$, this trend is reversed in the distilled case. We compute the $^{52} \mathrm{Cr} / ^{53} \mathrm{Cr}$ ratio to be 2300 (core) versus 5100 (homogeneous core).

We now return to our comparison with the prior work of \citet{bravoTypeIaSupernovae2024}. This work provides a detailed examination of the effects of chemical segregation on the characteristics of SNe\,Ia for several different progenitor types, including CONe, COFe, and CONeFe WDs. The results relating to chemically-segregated $^{56} \mathrm{Fe}$ cannot be directly compared to our results. However, there is also significant discussion of the effects of $^{22} \mathrm{Ne}$ distillation. The ``Full'' and ``Shell'' $1.06$\,\msol\ WD models correspond most closely to our core-distilled and shell-distilled models. Notably, the higher mass of the \citet{bravoTypeIaSupernovae2024} models results in a significantly higher central density -- $5.9 \times 10^7$\,\gcmcubed for the Full model, and $5.3 \times 10^7$\gcmcubed\, for the shell model. This is also reflected in a significantly higher yield of $^{56} \mathrm{Ni}$ -- $0.635$\, and $0.682$\,\msol\, respectively. Another difference is that \citet{bravoTypeIaSupernovae2024} conduct their simulations using a 1D hydrodynamics and nucleosynthesis code \citep[see][]{bravoSNRcalibratedTypeIa2019}.

Similarly to our results, \citet{bravoTypeIaSupernovae2024} find that $^{22} \mathrm{Ne}$ distillation has the potential to change the inferred character of the progenitor according to standard measures (i.e. the stable nickel profile and the presence of a central hole in the distribution of $^{56} \mathrm{Ni}$). The $^{58} \mathrm{Ni}$ yield of the Bravo-Full and Bravo-Shell models is $6.36 \times 10^{-2}$ and $1.30 \times 10^{-2}$\,\msol\ respectively. We find $7.17 \times 10^{-2}$ and $0.783 \times 10^{-2}$ for our core- and shell-distilled models. These values are somewhat similar and show the same general trend, however, they may not be directly comparable due to the much larger differences in $^{56} \mathrm{Ni}$ yield.

The authors further find that in sub-$M_{ch}$ explosions, the cobalt and copper abundances increase while manganese and several isotopes of chromium decrease. When comparing the ``Full'' model with its homogeneous equivalent, \citet{bravoTypeIaSupernovae2024} report lower yields of $^{50, 53, 54} \mathrm{Cr}$ (decreasing by a factor ten, factor four, and 60\% respectively). Conversely, $^{52} \mathrm{Cr}$ experiences an increase in abundance of 30\%. We find similar results for these isotopes when comparing $M_c$ with $M_{hc}$ (see Figure~\ref{fig:chromium}). In summary, we can confirm that the results of \citet{bravoTypeIaSupernovae2024} are overall consistent with our findings. 

\subsection{Velocity-Space abundance structure} \label{sec:vel_space}

\begin{figure*}
    \centering
    \includegraphics[width=0.95\linewidth]{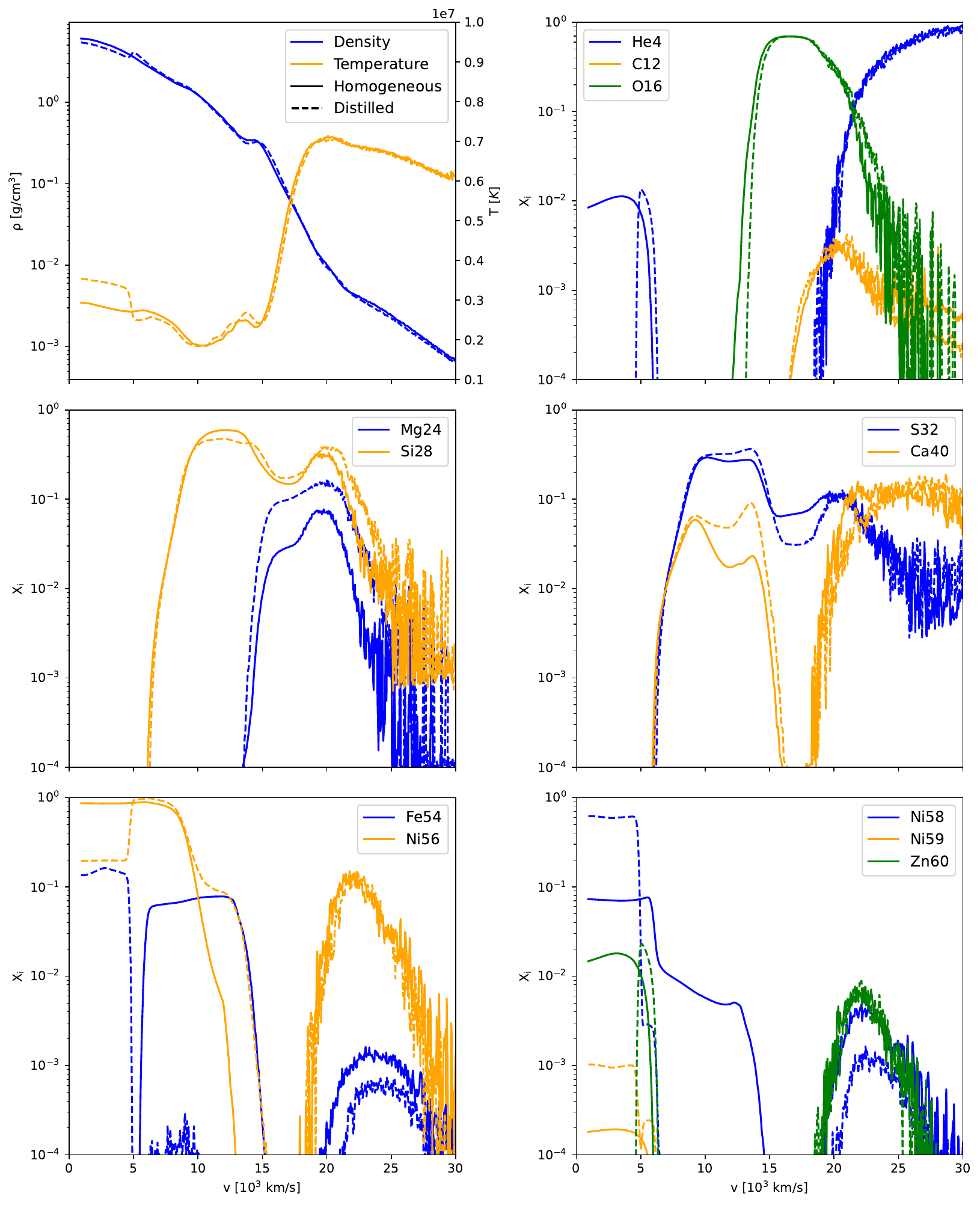}
    \caption{Velocity-space distribution of key isotopes in the 384 species network approximately $50$\, seconds after detonation, comparing the $M_c$ model with its homogeneous equivalent. Several isotopes -- including $^{28} \mathrm{Si}$, $^{32} \mathrm{S}$, $^{40} \mathrm{Ca}$, and $^{56} \mathrm{Ni}$ -- show multi-peaked velocity distributions, with deep minima at $v \sim 15{,}000 - 18{,}000$ \, \kmpersec. This trough separates the IME products produced at higher density in the core from the IME nucleosynthesis at lower density in the helium-rich portion of the ejecta (see Section~\ref{sec:vel_space}). The homogeneous and distilled distributions show significant differences at low velocities, particularly as relates to $^{54} \mathrm{Fe}$ and $^{58} \mathrm{Ni}$. This is due to the concentration of excess neutrons in the core region, leading to enhanced production of iron-group elements at the expense of radioactive nickel.}
    \label{fig:vel_space_core}
\end{figure*}

\begin{figure*}
    \centering
    \includegraphics[width=0.95\linewidth]{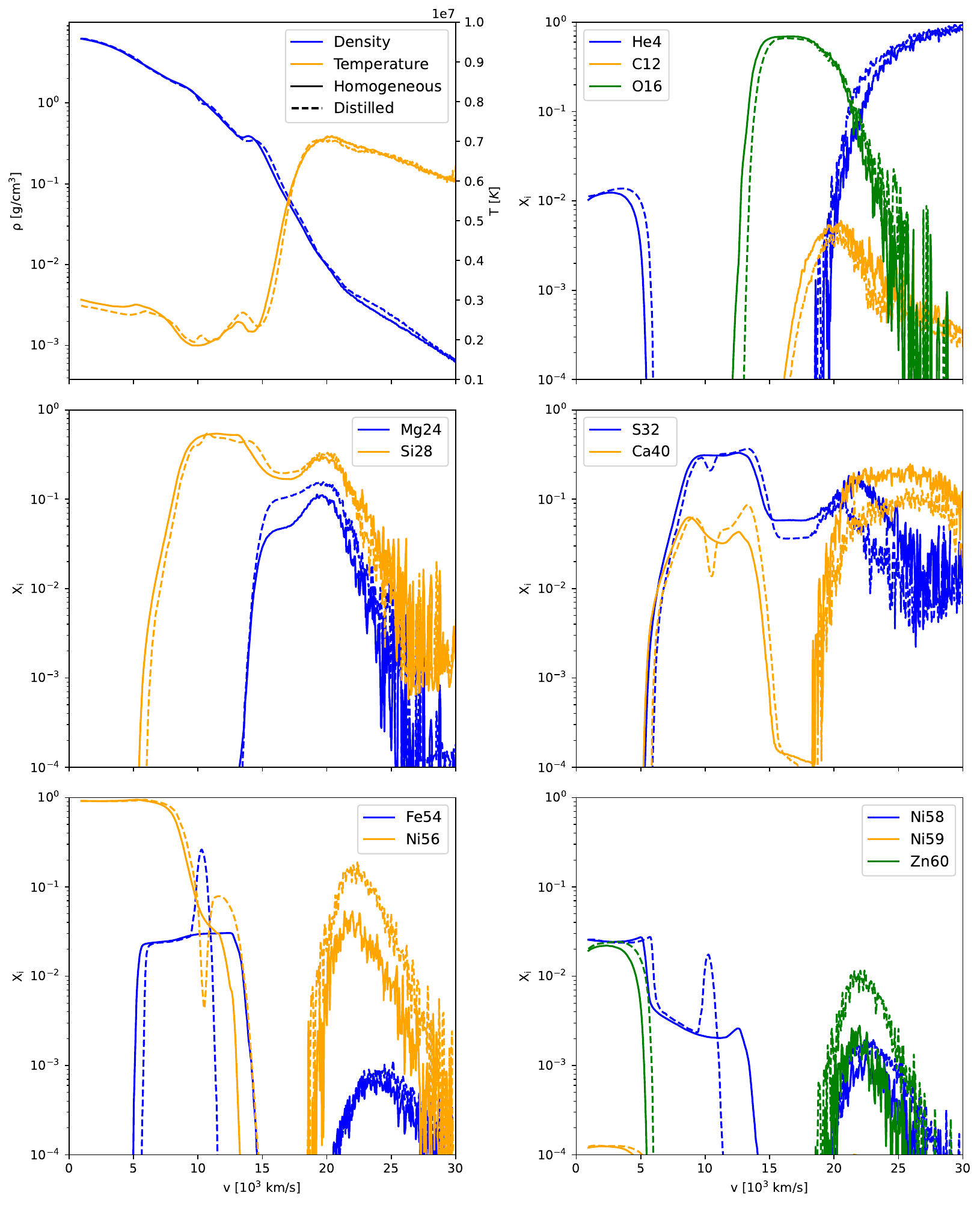}
    \caption{Velocity-space distribution of key isotopes in the 384 species network approximately $50$\, seconds after detonation, comparing the $M_s$ model with its homogeneous equivalent. As in Figure~\ref{fig:vel_space_core}, there are multi-peaked velocity distributions as a result of the onset of the helium-rich portion of the ejecta at $v \sim 15{,}000 - 18{,}000$ \, \kmpersec. The distilled distribution also shows peaks in $^{54} \mathrm{Fe}$ and $^{58} \mathrm{Ni}$ abundance at $v \sim 10{,}000$ \, \kmpersec, corresponding to the presence of the $^{22} \mathrm{Ne}$-rich shell.}
    \label{fig:vel_space_shell}
\end{figure*}

Figures~\ref{fig:vel_space_core} and \ref{fig:vel_space_shell} show the velocity-space abundance distributions of selected isotopes for the $M_c$ and $M_s$ models, compared to their homogeneous equivalents. Several species exhibit a double-peaked or multi-modal structure, with a pronounced abundance minimum at velocities $v \sim 15{,}000$–$18{,}000$\,\kmpersec. This affects species such as $^{28} \mathrm{Si}$, $^{32} \mathrm{S}$, $^{40} \mathrm{Ca}$, and $^{56} \mathrm{Ni}$ and occurs both in the homogeneous and distilled cases. The cause is the lower boundary of the helium-rich portion of the ejecta, which is apparent in the top-right panel of both figures as an increasing mass fraction  of $^{4} \mathrm{He}$ for $v \gtrsim 18{,}000$\,\kmpersec. At the intermediate densities found between $v \sim 15{,}000$–$18{,}000$\,\kmpersec, the C/O detonation is incapable of producing these intermediate mass elements. However, at lower densities, they can still be produced as products of the helium detonation.

The $M_s$ and $M_{hs}$ models show some differences in element distribution. The most pronounced feature is visible at $v \sim 10{,}000$ \, \kmpersec, where the distilled model shows a sudden decrease in $^{56} \mathrm{Ni}$, $^{32} \mathrm{S}$, and $^{40} \mathrm{Ca}$ along with a corresponding increase in $^{54} \mathrm{Fe}$ and $^{58} \mathrm{Ni}$. These peaked features correspond to tracers that originated in the thin shell. We also notice that there is a sudden drop-off in the abundance of these peaked elements in the distilled case, while there is a more gradual decline in the homogeneous case. This feature likely corresponds to the lack of $^{22} \mathrm{Ne}$ at radii outside the shell in the case of the distilled model. 

The $M_c$ and $M_{hc}$ models show a significant difference in the presence of $^{56} \mathrm{Ni}$ and $^{54} \mathrm{Fe}$ at low velocities. Whereas the homogeneous model shows a continuously high mass fraction of radioactive nickel through to very low velocities, the distilled model shows a sharp drop in $^{56} \mathrm{Ni}$ which is replaced by the presence of $^{54} \mathrm{Fe}$ and $^{58} \mathrm{Ni}$. In the homogeneous model, meanwhile, $^{54} \mathrm{Fe}$ is only found in significant amounts between $v \sim 5{,}000 - 15{,}000$ \, \kmpersec. This bears a similarity to ``abundance tomography'' studies from \citet{stehleAbundanceStratificationType2005, stehleAbundanceTomographyType2005}. These studies examined SN Ia 2002bo via this new method - the abundance distribution of the ejecta was derived by fitting a series of spectra obtained at close time intervals. These spectra were obtained via a Monte Carlo code that was modified to include abundance stratification. The abundance distribution graphs for 2002bo show a sharp drop-off of $^{56} \mathrm{Ni}$ abundance at $v \sim 3{,}000$ \, \kmpersec, with the remaining relative abundance made up by iron-group elements \citep[][Fig. 5]{stehleAbundanceStratificationType2005}. As discussed in Section~\ref{sec:yield_comparison}, \citet{bravoTypeIaSupernovae2024} also observe a central hole in the distribution of radioactive $^{56} \mathrm{Ni}$, which may influence the inferred character of the progenitor.

\begin{figure*}
    \centering
    \includegraphics[width=\linewidth]{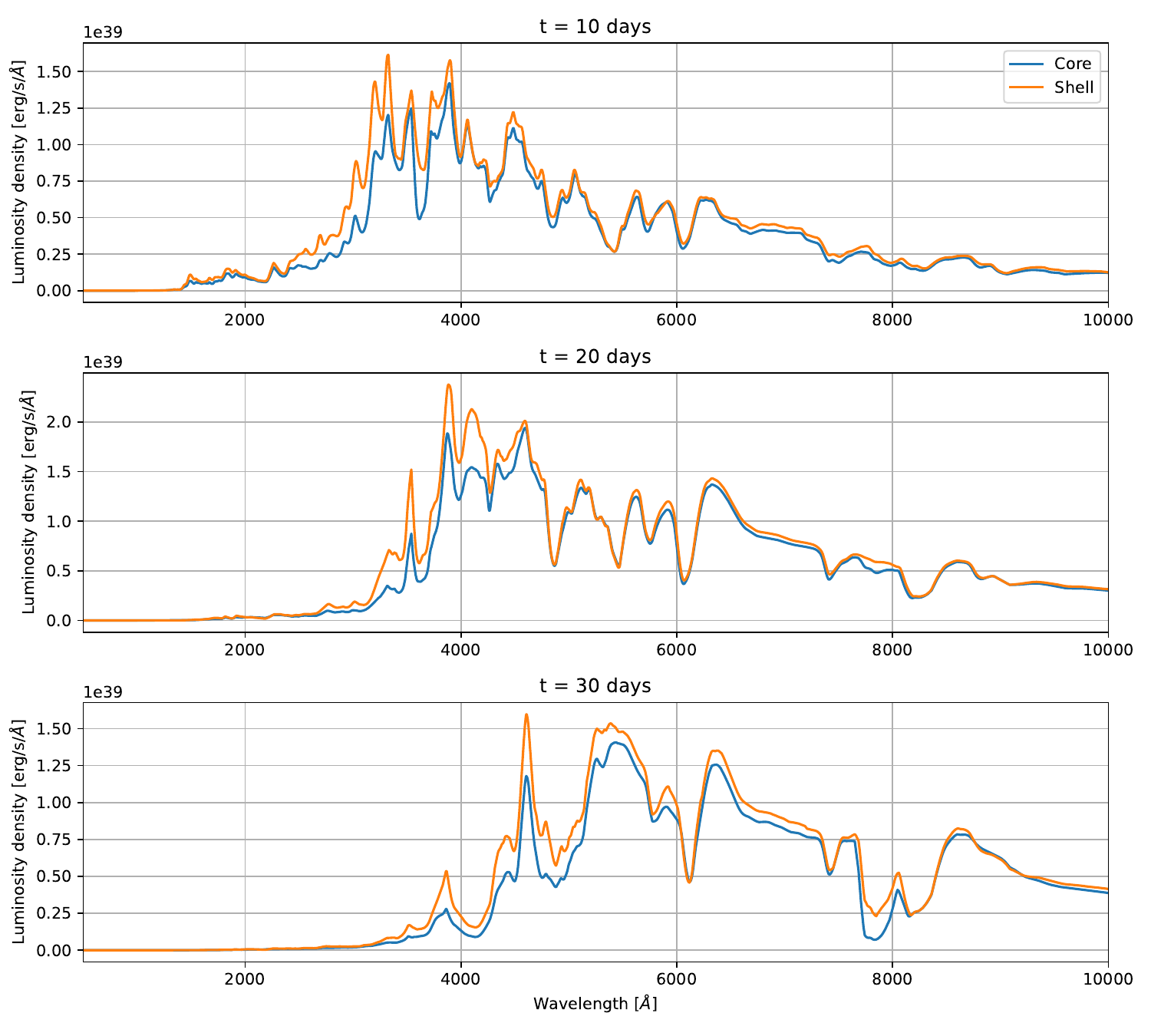}
    \caption{Comparison of spectra between the core and shell distilled models at various times after detonation, roughly corresponding to an early observation, peak luminosity and late-time observation. Beyond 40 days the spectrum becomes dominated by forbidden lines, which cannot be simulated using our existing \textsc{TARDIS} toolchain.}
    \label{fig:spectra}
\end{figure*}

\subsection{Synthetic emerging spectra}

To model the emergent spectra from our explosion models, we employ the Monte Carlo radiative transfer code \textsc{TARDIS} \citep{kerzendorfSpectralSynthesisCode2014}. \textsc{TARDIS} simulates the propagation of photon packets through homologously expanding supernova ejecta, assuming radiative equilibrium and adopting the sharp photosphere approximation, i.e., the inner boundary of the emitting region is treated as a distinct, well-defined surface. This approximation is reasonable at early times, when the inner regions of the supernova remain optically thick and the observed light predominantly originates from a narrow region near optical depth $\tau \sim 1$. As such, \textsc{TARDIS} is particularly well-suited for generating synthetic spectra of SNe\,Ia during the photospheric phase (up to $\sim 40$ \, days post-explosion), when the ejecta are still largely opaque. Because \textsc{TARDIS} does not account for non-local thermodynamic equilibrium (non-LTE) effects, time-dependent ionisation, or nebular-phase emission, it is unsuitable for epochs later than $\sim$40\,d, when the ejecta become optically thin and the spectrum becomes dominated by recombination and forbidden-line emission.

We construct synthetic spectra using \textsc{TARDIS} at 10, 20, and 30 days after explosion for our four WD types. We show the ``formal integral'' spectrum of $M_c$ and $M_s$ in Figure~\ref{fig:spectra}. The formal integral is a method of generating the spectrum implemented in \textsc{TARDIS} which eliminates much of the Monte Carlo noise. These epochs are chosen to sample key stages of the light curve: the early rise (pre-maximum), peak luminosity, and the transitional phase leading to the nebular stage. We have chosen not to show spectra for our homogeneous models as we are primarily interested in whether long and short cooling delay WDs might be distinguished by their spectral features. Thus, we primarily compare our distilled models.

We must provide a target luminosity value for \textsc{TARDIS} to match at each of these three epochs -- we estimated this using energy production estimates derived from the $^{56} \mathrm{Ni} \rightarrow ^{56} \mathrm{Co} \rightarrow ^{56} \mathrm{Fe}$ decay chain. This approach was first articulated by \citet{arnettTypeSupernovaeAnalytic1982} and later expanded upon by others to include the effects of positron heating and gamma-ray leakage \citep[see e.g.,][]{nadyozhinPropertiesNICO1994, chatzopoulosModelingLightCurve2009, valentiBroadlinedTypeIc2007}. We use the generalised semi-analytical model described in \citet{chatzopoulosGeneralizedSemiAnalyticalModels2012} for a radioactive-decay-powered light curve with ejecta in homologous expansion. This model requires the target time and initial $^{56} \mathrm{Ni}$ mass, which was taken from Table~\ref{tab:ejecta_unstable}. Additionally, we provide the initial mass of the progenitor, the ejecta velocity, and a gamma-ray leakage factor. We use a progenitor radius of $R_0 = 5 \times 10^8$\,\cm, an ejecta velocity of $v = 30{,}000$\,\kmpersec, and a gamma-ray leakage factor which ensured almost all gamma rays and positrons are trapped.

Lastly, we must express the state of our ejecta as an input to \textsc{TARDIS}. Due to the highly symmetrical nature of our detonation, we have no preferred direction or features to capture when converting our ejecta into a one-dimensional profile. Thus, we sample our ejecta by taking a spherical profile of the three-dimensional cloud of tracer particles, keeping those between $0.5 \times 10^{11} < r < 1.5 \times 10^{11}$\ \cm\, of the centre of the simulated box. The averaged velocity, density, and chemical composition along this profile are then used as the input to \textsc{TARDIS} in the form of a CSVY file. 

\subsection{Comparison of SDEC Spectra} \label{sec:sdec}

\begin{figure*}
    \centering
    \includegraphics[width=\linewidth]{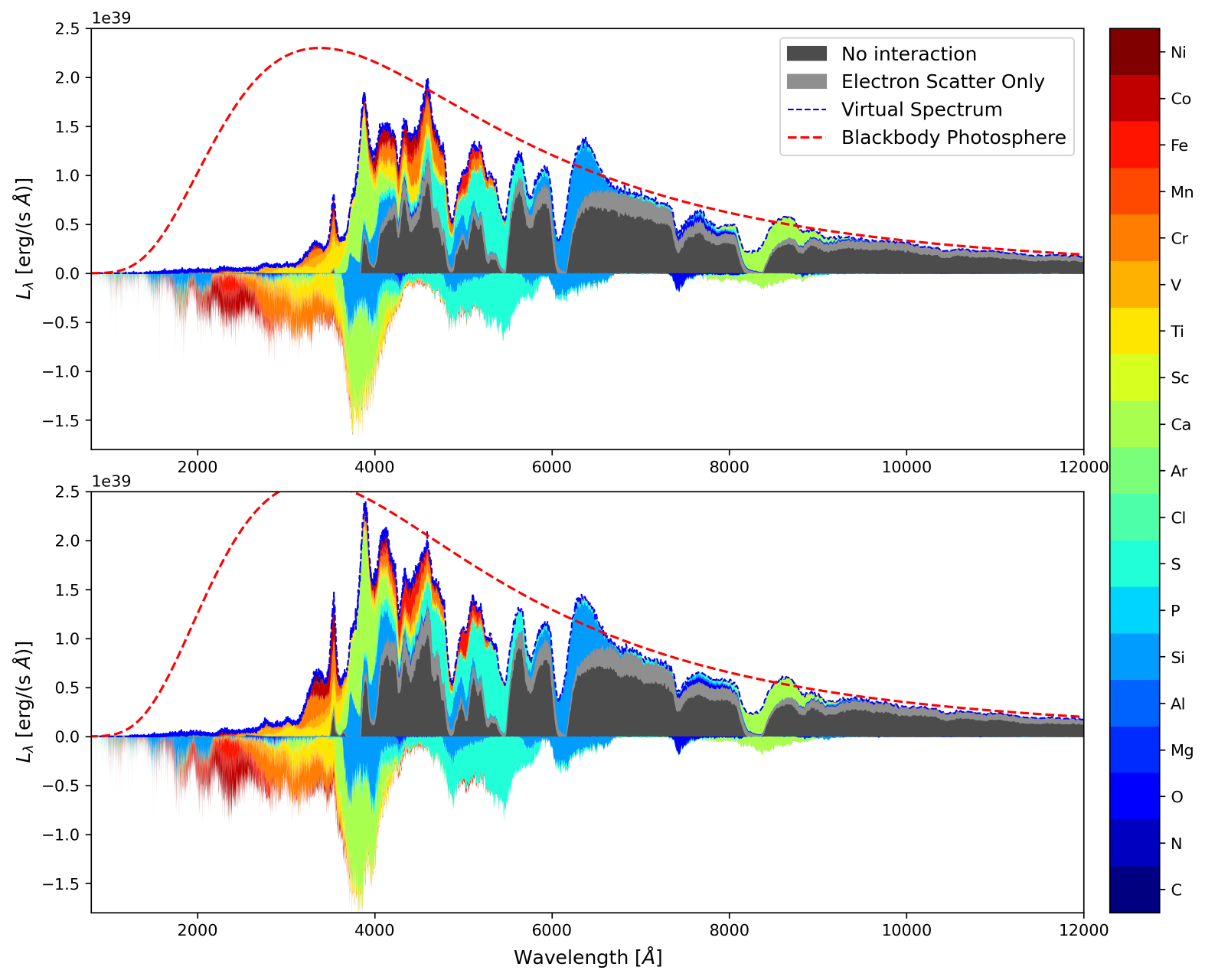}
    \caption{Spectral element decomposition for the $M_c$ (top) and $M_s$ (bottom) models at $t = 20$\,days. See Section~\ref{sec:sdec} for further information.}
    \label{fig:sdec}
\end{figure*}

To aid in the interpretation of the resulting spectra, we employ Spectral element DEComposition (SDEC) plots. These are post-processing diagnostics that decompose the emergent flux into contributions from individual ions by tracking the last bound-bound interaction each escaping photon packet experienced, both in emission and absorption \citep[See also Figure 6 of ][for further explanation and comparison]{kromer3DDeflagrationSimulations2013}. The SDEC analysis is based on the same Monte Carlo simulation as the synthetic spectrum but provides additional insight into some of the absorption and emission features. In the SDEC plot shown in Figure~\ref{fig:sdec}, positive flux components trace re-emission by specific ions, while negative flux components indicate line absorption (i.e., line blanketing), either by absorption or scattering. SDEC plots are particularly useful for identifying which ions and atomic transitions are responsible for specific features in the synthetic spectrum. They also help in diagnosing the physical conditions of the ejecta -- such as ionisation state, temperature structure, and the depth of line formation by highlighting how different ions absorb and re-emit radiation across the wavelength range. However, they do not represent observable fluxes directly. Their interpretation can become ambiguous in the presence of multiple scattering events or when several overlapping lines contribute to the same spectral feature. The SDEC plot for $t = 20$\, days is shown in Figure~\ref{fig:sdec}, but the corresponding SDEC plots for the other times are not included as they do not add much to our understanding.

When comparing the SDEC plots of the $M_c$ and $M_s$ models at 10, 20, and 30 days, the synthetic spectra produced by \textsc{TARDIS} exhibit broadly similar features across all epochs. Slight differences in total luminosity are evident, with the $M_s$ model appearing modestly brighter at early times, probably caused by differences in ejecta composition and density structure that influence radiative diffusion. Although the $M_c$ model is expected to synthesise a larger mass of centrally concentrated iron-group elements (e.g., $^{58} \mathrm{Ni}$ and $^{59} \mathrm{Ni}$), and the $M_s$ model exhibits more intermediate-mass elements such as Si and Ca in its outer layers, these differences in composition do not result in strong spectral differences in the epochs considered here. 

The most prominent absorption and re-emission features in the synthetic spectra include:
\begin{enumerate}
    \item \textbf{At 10 and 20\, days:} Broad \ion{Fe}{II} and \ion{Co}{II} blends contribute to significant line blanketing in the ultraviolet and blue optical regions (2500--5000\,\AA), with stronger suppression in the $M_c$ model, consistent with its enhanced iron-group content. 
    \item \textbf{At 20\,days:} The \ion{Si}{II} $\lambda$6355 line produces strong absorption near 6100--6200\,\AA, which is a distinctive feature of SNe\,Ia during the photospheric phase, clearly visible in both $M_c$ and $M_s$ \citep[see e.g.,][]{zhaoStudySiII2021}.    
    \item \textbf{At 30\,days:} The \ion{Ca}{II} near-infrared triplet is responsible for broad absorption around 8000\,\AA\ and a re-emission peak near 8800 -- 9200\,\AA. This feature grows in strength as the ejecta cools and expands.
\end{enumerate}

These features are very similar in both models, with only slight differences in strength and width. The $M_c$ model exhibits slightly deeper blue/UV blanketing, consistent with its enhanced production of stable iron-group isotopes such as $^{58} \mathrm{Ni}$ and $^{59} \mathrm{Ni}$, which dominate the opacity in this wavelength range. Although $M_s$ contains significantly more $^{55} \mathrm{Co}$, this isotope alone does not account for early-time line blanketing, as it contributes less to persistent opacity than the more stable Fe-group species prevalent in $M_c$. The $M_s$ model shows marginally stronger \ion{Si}{II} and \ion{Ca}{II} absorption, consistent with enhanced IME content in its outer ejecta layers.

\begin{figure}
    \centering
    \includegraphics[width=\linewidth]{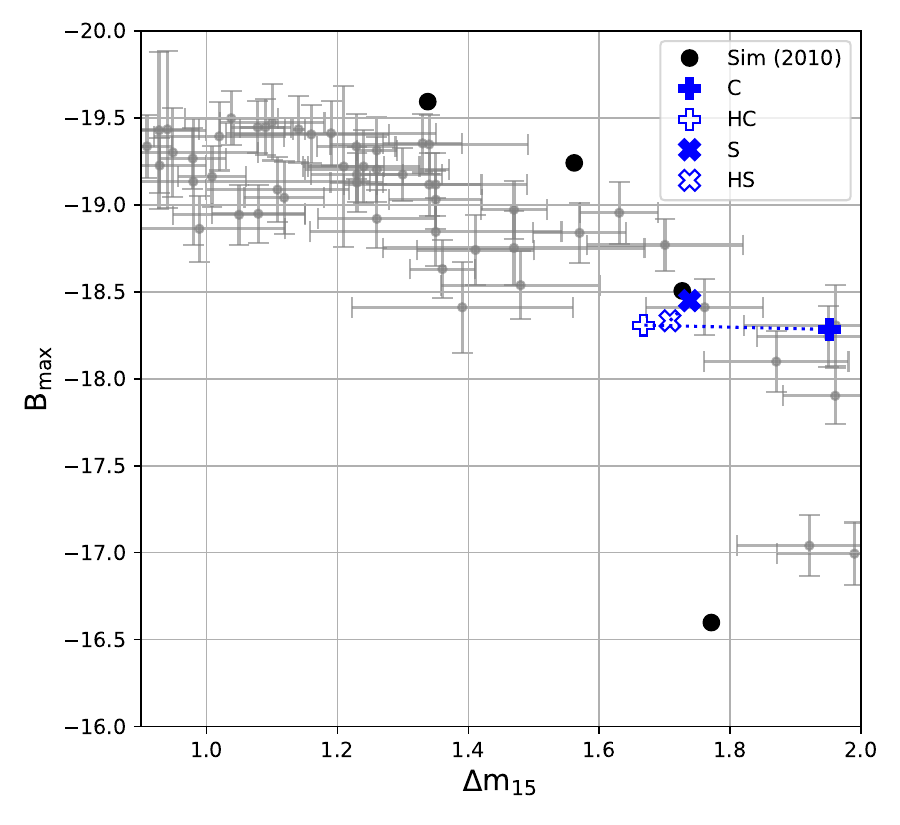}
    \caption{B-band light curve peak magnitude ($B_{max}$) versus the decline-rate parameter ($\Delta m_{15}$), partially reproduced from \citet{simDetonationsSubChandrasekharmassC+O2010}. The solid circles correspond to numerical simulations of WD detonations while the grey crosses are observations of SNe Ia with distance modulus $\mu > 33$. Our four models have been over-plotted in blue, with the corresponding distilled and homogeneous models connected by a dotted line.}
    \label{fig:sim_wlr}
\end{figure}

These results highlight a key limitation of early-time spectra as diagnostics for distinguishing between neutron-rich ejecta configurations. While the underlying nucleosynthetic differences between the $M_c$ and $M_s$ models are significant, their imprint on the photospheric-phase spectra is modest. At nebular times (e.g., $\sim 300$ days after explosion), the ejecta become optically thin and the observed spectra are dominated by forbidden lines of iron-group elements, particularly \ion{Fe}{II}, \ion{Fe}{III}, and \ion{Co}{III}. These late-time spectra are especially sensitive to the distribution of stable neutron-rich isotopes and may provide more robust observational discriminants between the $M_c$ and $M_s$ models than photospheric-phase spectra. These lines originate from the inner, denser regions where radioactive decay products deposit their energy. As such, late-time spectra are sensitive to the distribution and composition of the inner ejecta and offer a promising diagnostic for distinguishing between the $M_c$ and $M_s$ models. In the $M_c$ model, the centrally concentrated, neutron-rich core leads to stronger emission from stable iron-group isotopes such as $^{58} \mathrm{Ni}$ and $^{54} \mathrm{Fe}$, potentially manifesting as enhanced [\ion{Ni}{II}] and [\ion{Fe}{II}] emission features, particularly in the near-infrared (e.g., [\ion{Ni}{II}] $\lambda$7378). 

We can also use our spectra to estimate the B-band light curve peak magnitude $B_{max}$ and decline-rate parameter $\Delta m_{15}$ for each of our models. This parameter is defined as the change in B-band magnitude between maximum light and an epoch 15 days later providing a useful comparison point with observations and in particular the Width-Luminosity Relation. For our purposes, we take the $t = 20$\,days to be the B-band maximum time and compute an additional spectrum at $t = 35$\,days to find $\Delta m_{15}$. We use the Bessel B passband filter \citep{bessellUBVRIPassbands1990} and compute an integral over the spectrum output by \textsc{TARDIS}. We find the following magnitudes for $(\Delta m_{15}, B_{max})$ -- $M_c$: $(1.95, -18.28)$, $M_{s}$: $(1.74, -18.45)$, $M_{hc}$: $(1.67, -18.31)$, and $M_{hs}$: $(1.71, -18.33)$. Comparing with Figure~\ref{fig:sim_wlr}, we can see that the shell and homogeneous models all produce a B-band magnitude and decline rate very similar to the $0.97$\,\msol C/O WD model. It would thus be difficult to distinguish between the two. 

While the lowest-mass bare detonation model from \citet[][rightmost black point]{simDetonationsSubChandrasekharmassC+O2010} predicts a fast-declining event that appears very faint in B-band and lies below the observed distribution of SNe Ia, the core-distilled model predicts a very fast-declining event that occupies the reasonably bright ($-18.5 < B_{max} < -18.0$), yet fast declining ($1.9 < \Delta m_{15} < 2.0$), region of the parameter space, consistent with observed events. The core-distilled model is thus capable of reproducing these fast-declining events, presenting another possibly distinguishing feature in observations. One test of this would be to determine if such fast-declining but reasonably bright events occur only in old stellar populations where such long cooling delays have had time to elapse. 

\section{Conclusions} \label{sec:conclusion}

We have investigated the impact of $^{22}\mathrm{Ne}$ distillation on the thermonuclear explosions of $1.0$\,\msol\ WDs. Using the moving-mesh \textsc{AREPO} code, we were able to create 3D structures corresponding to the 1D chemical profiles of a $^{22} \mathrm{Ne}$-enriched core WD ($M_c$) and $^{22} \mathrm{Ne}$-enriched shell WD ($M_s$) from \citet{blouin22NePhaseSeparation2021}. We also created chemically-homogeneous WDs which acted as comparison to these distilled models ($M_{hc}$ and $M_{hs}$). We artificially detonated these WDs and followed their evolution through to homologous expansion using \textsc{AREPO}. We employed the \textsc{YANN} nuclear network code and Lagrangian tracers for post-processing, upscaling the 55-species network used for the energy release in \textsc{AREPO} to detailed 384-species yields. Lastly, we used the spatial, velocity, and chemical structure obtained from \textsc{AREPO} and used it as an input to the radiative transfer code \textsc{TARDIS}. Our goal was to assess whether the altered internal composition resulting from crystallisation and chemical separation leaves observable signatures in the explosion ejecta and emergent spectra.

We find that the two distilled models yield comparable $^{56} \mathrm{Ni}$ masses ($0.40$ and $0.45$\,\msol) and light element abundances, consistent with normal to moderately luminous SNe\,Ia. However, the $M_c$ model exhibits enhanced production of stable neutron-rich isotopes such as $^{58} \mathrm{Ni}$ and $^{59} \mathrm{Ni}$, concentrated in the inner ejecta, while the $M_s$ model retains more $^{55} \mathrm{Co}$ and $^{54} \mathrm{Fe}$ in its middle layers. These differences arise from the distinct spatial distribution of neutron excess caused by $^{22} \mathrm{Ne}$ distillation prior to explosion, but early-time synthetic spectra (up to 40 days post-explosion) show only modest differences. Line blanketing from \ion{Fe}{II} and \ion{Co}{II} is marginally stronger in $M_c$ due to its higher stable Fe-group content, while $M_s$ shows slightly deeper \ion{Si}{II} and \ion{Ca}{II} features, consistent with its IME-rich middle layers. However, these differences are not sufficient to distinguish between the models based on early photospheric spectra alone. We suggest a further investigation of these results should include radiative transfer calculations concentrating on the forbidden line transitions during the nebular phase.

We find more significant differences when we compare the yields of the distilled and homogeneous models. The distilled models produce significant enhancement of $^{15} \mathrm{N}$ in comparison to the homogeneous case, with a $^{15} \mathrm{N} / ^{14} \mathrm{N}$ ratio of 12.0 (shell) and 10.7 (core). When comparing the abundance ratio of most other elements, the yield of the shell distilled model is reduced. This result may be interesting for research in dust and planetary formation, but we expect that it would be challenging to observe due to the available ionisation states of \ion{N}{I}. The $M_c$ model produces $7.2 \times 10^{-2}$\,\msol\, in comparison to a yield of $2.0 \times 10^{-2}$\,\msol\, in the homogeneous case, as a result of the concentration of neutron-rich material towards the high-density core. This has a significant impact on observations and may complicate the use of stable nickel mass as a discriminator between sub-$M_{ch}$ and $M_{ch}$ detonations. Distillation also had the effect of significantly increasing the number abundance of Ca, and Ar, while reducing the abundance of Mn.

Our results underscore the importance of considering distillation-driven composition gradients when modelling thermonuclear explosions of massive WDs. Although early spectra remain largely degenerate between the core- and shell-enriched configurations, nebular-phase observations, when radioactive decay products dominate, are expected to offer clearer diagnostics of the neutron-rich isotope distribution. Lastly, we also observed a higher B-band decline rate for the core distilled model compared to the shell-distilled model and its homogeneous counterparts. This might be another factor in distinguishing distilled from homogeneous WDs in observations. Distillation may then also provide an explanation for observed SNe which boast high decline rates but are nonetheless relatively bright -- this could be tested by determining if such events are found in old stellar populations only.

\section*{Code Availability}

The \textsc{AREPO} code is publicly available \href{https://arepo-code.org}{here}, however, some features are only available in the private ``Development'' branch. Access to the development branch is subject to approval from \textsc{AREPO}'s development team.

\section*{Software}

This work made use of the following open-source software projects: NumPy \citep{harrisArrayProgrammingNumPy2020}, MatPlotLib \citep{hunterMatplotlib2DGraphics2007}, SciPy \citep{virtanenSciPy10Fundamental2020}. This research made use of \textsc{tardis}, a community-developed software package for spectral synthesis in supernovae \citep{kerzendorfSpectralSynthesisCode2014}. The development of \textsc{tardis} received support from GitHub, the Google Summer of Code initiative, and from ESA's Summer of Code in Space program. \textsc{tardis} is a fiscally sponsored project of NumFOCUS. \textsc{tardis} makes extensive use of astropy and PyNE.

\section*{Acknowledgements}

This research was undertaken with the assistance of resources and services from the National Computational Infrastructure (NCI), which is supported by the Australian Government. Computation time was contributed by the Australian National University through the Merit Allocation Scheme (ANUMAS), and the National Computational Merit Allocation Scheme (NCMAS). U.P.B was supported by the Australian Government Research Training Program (AGRTP) Fee Offset Scholarship and ANU PhD Scholarship (Domestic). This research was supported in part by the National Science Foundation grant PHY-1748958 to the Kavli Institute for Theoretical Physics (KITP). IRS and SB thank Frank Timmes for stimulating discussions in November 2022 during the KITP workshop ``White Dwarfs from Physics to Astrophysics'' and the associated KITP Program: ``White Dwarfs as Probes of the Evolution of Planets, Stars, the Milky Way and the Expanding Universe.'' U.P.B and IRS offer sincere thanks to R\"udiger Pakmor for the use of his \textsc{YANN} code and for his assistance with the literature review on this paper. U.P.B and IRS also extend their gratitude to Wolfgang Kerzendorf and Stuart Sim for their technical expertise relating to the \textsc{TARDIS} code and helpful suggestions regarding the calculation of the B-band magnitudes and decline rates. 


\section*{Data Availability}

The data underlying this article will be shared on reasonable request to the corresponding author.


\bibliographystyle{mnras}
\bibliography{main}





\bsp	
\label{lastpage}
\end{document}